\documentclass[%
reprint,
amsmath,amssymb,
aps,
]{revtex4-2}

\usepackage{graphicx}
\usepackage{dcolumn}
\usepackage{bm}
\usepackage{chemformula}
\usepackage{amsmath,scalerel}
\usepackage{braket}
\usepackage{subfigure}
\usepackage{cleveref}
\usepackage{flushend}
\usepackage{verbatim}
\usepackage{xr}
\usepackage[T1]{fontenc}
\usepackage[utf8]{inputenc}
\usepackage[normalem, normalbf]{ulem}

\usepackage{changes}
\usepackage{cancel}
\usepackage{xcolor}
\makeatletter
\newcommand*{\addFileDependency}[1]{
	\typeout{(#1)}
	\@addtofilelist{#1}
	\IfFileExists{#1}{}{\typeout{No file #1.}}
}
\makeatother

\begin{document}
	
	\preprint{APS/123-QED}
	
	\title{Role of \textit{host/guest} coupling in stabilizing the phases of the over-tolerant hybrid perovskite \ch{MHyPbX3}}
	
	\author
	{Pradhi Srivastava$^{1}$, Sayan Maity$^{2}$, Varadharajan Srinivasan$^{2}$}
	\affiliation{$^{1}$Department of Physics, Indian Institute of Science Education and Research Bhopal, Bhopal 462 066, India}
	\affiliation{$^{2}$Department of Chemistry, Indian Institute of Science Education and Research Bhopal, Bhopal 462 066, India}
	\date{\today}

	\begin{abstract}
		\ch{MHyPbBr_xCl_{3-x}} exhibits an interesting temperature-induced phase diagram transitioning from monoclinic (Phase-II) to orthorhombic (Phase-I) and eventually to the high-temperature disordered cubic phase. However, experimental observations indicate the absence of either the cubic or orthorhombic phases in compositions with x=0 (chloride) and x=3 (bromide), respectively. We explain the composition dependence of the phase transition sequence in \ch{MHyPbX3} (X=Cl, Br) from the perspective of a \textit{host/guest} framework for the system. We argue that the sequence of phase transitions in \ch{MHyPbX3} can be anticipated based on the competition between \textit{host} distortion and \textit{host/guest} coupling energies. A dominant coupling in \ch{MHyPbCl3} ensures the stabilization of the intermediate Phase-I while pushing the transition to cubic phase beyond its decomposition temperature. On the other hand, a balance of the energy scales in \ch{MHyPbBr3} suppresses Phase-I and stabilizes the cubic phase. Our expectations were supported by first-principles based estimates of the transition temperatures in both compounds, which yield trends in agreement with experimental observations. Furthermore, {\it ab initio} molecular dynamics simulations revealed that the cubic phase of \ch{MHyPbBr3} results from a disorder over locally orthorhombic structures, a clear manifestation of the balance of energy scales. We propose that the same concept can be employed to predict the stable phases in other hybrid perovskites.

	\end{abstract}

	\maketitle

	\section{\label{sec:intro}INTRODUCTION}
	Organic-inorganic lead halide perovskites (OIHPs), with the formula \ch{APbX3} (X=Cl, Br, I), comprise an organic \textit{guest} molecule at A-site encompassed by the inorganic PbX$_6$ octahedra that act as the \textit{host}. The broad spectrum of realizable chemical compositions, including variations in the size and shape of the organic component, is one of the primary reasons for the remarkable versatility observed in hybrid perovskites, allowing for extensive control and fine-tuning of their properties.~\cite{stranks2015metal,saparov2016organic} Given the immense technological potential of OIHPs, a comprehensive study of their structure-property relationship has to accompany the discovery of new materials. In particular, this would help identify stable phases as well as overcome challenges in their chemical stability and environmental impact~\cite{babayigit2016ethirajan}. 
	
\begin{figure}[tbh!]
		\centering
		\includegraphics[width=0.4\textwidth]{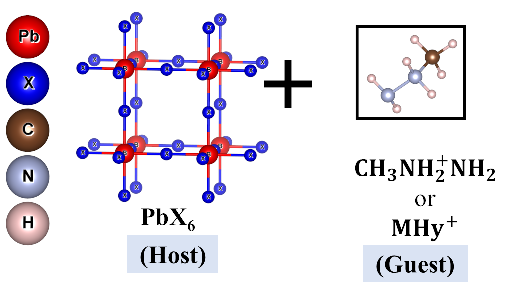}
		\caption{\label{fig:comp_phase_dig}~\textit{Host/guest} (HG) pair considered in this work. Lead-halide octahedra (\ch{PbX6}) make up the \textit{host}, and the \ch{MHy} is the \textit{guest}.}
	\end{figure}

	Recently, M\c{a}czka, {\it et al.}, successfully synthesized an over-tolerant member of 3D OIHPs, employing a relatively large \textit{guest}, methylhydrazinium (\ch{MHy+}) at its A-site (see Fig.~\ref{fig:comp_phase_dig}). The room temperature structures of \ch{MHyPbX3} (X = Br, Cl) exhibit a pronounced non-centrosymmetric/polar nature, unlike other members of the class. Temperature-dependent structural investigations using several experimental techniques revealed that the perovskites should undergo a final phase transition to the highly disordered and highest symmetric cubic phase~\cite{vogt1993high,dittrich2015temperature,herz2016charge}. However, for \ch{MHyPbCl3}, only one \textit{order} to \textit{order} transition -- from less polar (low-symmetry, Phase-II) to more polar phase (orthorhombic Phase-I) -- was seen at $342$ K. The compound decomposes at around $490$ K before any further transition~\cite{maczka2020three}. 
	On the other hand, \ch{MHyPbBr3}, was found to transition directly from an ordered polar (low-symmetry, Phase-II) to a disordered non-polar phase (cubic phase) at $420$ K. The intermediate Phase-I is found to be absent here in contrast to the observed phases in chloride analog~\cite{maczka2020methylhydrazinium}. 
	
	Interestingly, all three phases appear for the mixed halide \ch{MHyPbBr_xCl_{3-x}}, with the thermal stability range of Phase-I increasing with the chloride composition.~\cite{drozdowski2022three} This intriguing phase diagram motivated us to theoretically investigate the stability of different phases of \ch{MHyPbX3} (\ch{X=Cl,~Br}). 
	
	The sequence of phase transitions in a \textit{guest/host} coupled system can also be explained qualitatively by understanding the relative magnitudes of several interactions taking place in such a coupled system.~\cite{pradhi_mhpc,svane2017strong}.
	There are three major energy scales that can be identified in such systems, namely (a) strength of \textit{host} elasticity, denoted by $\Omega_S$, (b) \textit{host-guest} (HG) coupling strength manifested by hydrogen and co-ordinate bonds, denoted by $\lambda_{coup}$, and (c) dipole-dipole interaction between \textit{guest} species, denoted by $J_{dipole}$.~\cite{pradhi_mhpc} Consequently, the phase diagram can also be divided into regimes for strong and weak coupling, which could further indicate the driving force responsible for structural phase transitions under different scenarios. 
	Hence, estimating the relative magnitudes of different interactions can provide insights into the absence of phases in otherwise similar halide analogs. 
	
	Herein, we primarily focus on rationalizing the absence of the cubic phase in \ch{MHyPbCl3} and the orthorhombic phase in \ch{MHyPbBr3}. In particular, we compute the temperature-dependent relative Gibbs free energies of the two ordered phases (I and II) in each compound using a quasi-harmonic approximation to vibrational contributions. Phase-I has an unstable mode in both the experimentally synthesized phase of \ch{MHyPbCl3}~\cite{pradhi_mhpc} and the theoretically constructed phase of \ch{MHyPbBr3}. In both cases, dynamically stable structures separated by low energy barriers are obtained by displacement along the unstable mode. Hence, we model the contribution to Gibbs free energy of this unstable eigenmode using a quartic potential. This resulted in the determination of phase transition temperature in \ch{MHyPbCl3}, which is in good agreement with the experimental observations. Interestingly, on applying the same model to Phase-I of \ch{MHyPbBr3}, we find that it gets stabilized at around $430$ K, only slightly beyond the transition temperature of the experimental cubic phase (around $420$ K). The overlapping stability regions of the two phases prompted us to probe the nature of the cubic phase through \textit{ab initio} molecular dynamics simulations. Our analyses revealed that the cubic phase is disordered over several orthorhombic-like \textit{guest} ordered structures, which explains the close proximity of the two transition temperatures. The sequence of phase transition, as demonstrated from experiments, is also explained qualitatively by comparing the ratio of \textit{host/guest} coupling strength to the \textit{host}-elasticity for chloride and bromide perovskites. We also find that the unstable mode can be made stable upon applying a pressure of around $4$ GPa and $5$ GPa for Cl and Br perovskites, respectively, thereby suggesting a plausible way to stabilize the more polar structure at lower temperatures. Further, for the cubic phase, we compare the rotational energy barriers between specific \textit{guest}-orientations spanned by the \textit{guest}. This is done under the assumption that the \textit{guest} orientation in chloride perovskite would be very similar to that in the bromide perovskite, given the similarity in their low-temperature optimized structures. The calculated energy barrier for the former suggests a transition to cubic phase around $560$ K, which is much beyond its decomposition temperature, thereby explaining its absence in the experimental phase diagram.
	
	\section{\label{sec:comp_details}Computational Details}
	\subsection{Details of structural optimizations}
	Ground state density functional theory (DFT) computations in this work were carried out using the QUANTUM ESPRESSO (QE) code.~\cite{giannozzi2009quantum,giannozzi2017advanced} The ionic cores were modeled through the RRKJ-ultrasoft pseudopotential~\cite{rappe1990optimized,vanderbilt1990soft}. The valence electrons for different elements that were treated explicitly included $5d^{10}6s^{2}6p^2$ for Pb, $3s^2p^5$ and $4s^2p^5$ for Cl and Br, respectively, $2s^2 2p^3$ for N, $2s^{2}2p^{2}$ for C and $1s^1$ for H. Calculations were performed with the (PBEsol)~\cite{perdew1996generalized,perdew2008restoring} generalized gradient approximation to the exchange-correlation functional. The vdW interactions in the crystals were included through second-generation DFT-D2 scheme as proposed by Grimme~\cite{grimme2006semiempirical,barone2009role,lee2010higher} since they are required for capturing the accurate structural and dynamical properties of OIHPs~\cite{egger2014role,wang2014density,beck2019structure}. The kinetic energy cut off for the plane wave basis set was set to 80 Ry, and that for the augmented charge was represented by $480$ Ry and $500$ Ry for \ch{MHyPbCl3} and \ch{MHyPbBr3}, respectively, following convergence tests. Structural optimizations were performed using the BFGS algorithm with a force convergence criterion of $10^{-5}$ Ry/Bohr. The electronic Brillouin zone (BZ) of the system was sampled using a $4\times 2\times 2$ Monkhorst-Pack $k-$point mesh. Phonon dispersion and the zero point energy (ZPE) for different structures were calculated using the finite displacement approach~\cite{kresse1995ab,parlinski1997first,chaput2011phonon}, as implemented in PHONOPY~\cite{togo2008first,phonopy}, on a $2\times2\times2$ supercell and $2\times2\times2$ $k$-point mesh. The dispersion thus calculated was used to obtain thermodynamic properties~\cite{wallace1972thermodynamics}, such as vibrational entropies, free energies, and specific heat capacities under the harmonic approximation. 
	\subsection{Details of molecular dynamics simulations}
	{\it Ab initio} molecular dynamics (AIMD) simulations in the canonical (NVT) ensemble were performed in the $2\times2\times2$ supercell (112 atoms) with experimental cubic lattice parameters of \ch{MHyPbBr3} employing Born-Oppenheimer molecular dynamics (BOMD)~\cite{born1985quantentheorie,born2000quantum,barnett1993born}, as implemented in VASP~\cite{kresse1993ab,kresse1996efficiency,kresse1996efficient}, at 470 K (beyond the experimental phase transition temperature, $420$ K). A $k$-point mesh of $2\times2\times2$ was employed to sample the BZ. The MD used a time step of 0.5 fs and a Nos{\'e}-Hoover thermostat with a Nos{\'e} mass of 0.02 for controlling the temperature.~\cite{nose1984unified,hoover1985canonical,kresse1994ab}. At each time step, DFT calculations were performed using the projector-augmented wave (PAW) pseudopotential method~\cite{kresse1999ultrasoft} and PBEsol-D2 functional. A plane-wave kinetic-energy cutoff of 900 eV with an electronic energy convergence threshold of $10^{-9}$ eV was used. 
	The simulations were equilibrated for the first 5 ps, after which a production run of 37 ps (for frozen cage) and 70 ps (for flexible cage) were used for all analyses.
	
	\section{Results and discussion}
	
	\subsubsection{Optimization of different phases}
\begin{figure}[tbh!]
		\centering
		\includegraphics[width=0.4\textwidth]{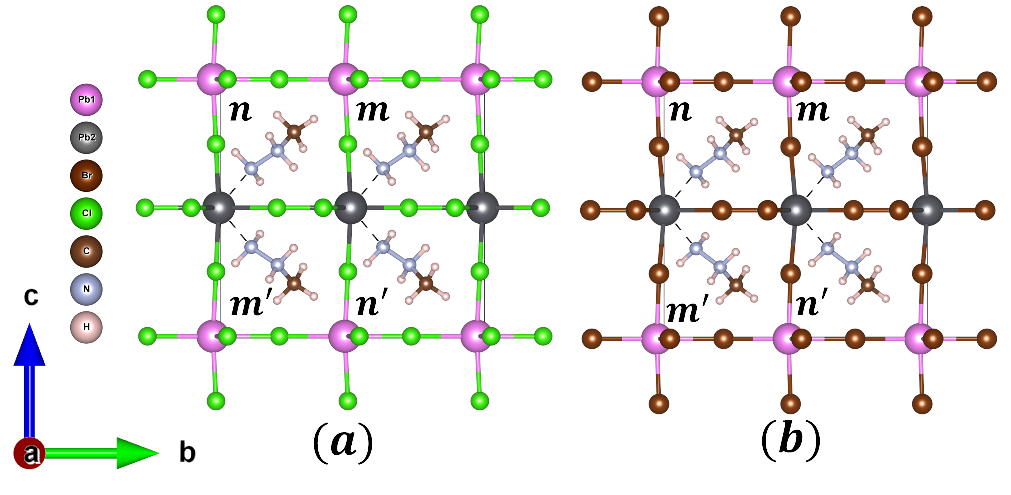}
		\caption{\label{fig_ch5:mhpb_mhpc_ht}~Optimized Phase-I of (a) \ch{MHyPbCl3} and (b) \ch{MHyPbBr3}. Phase-I of \ch{MHyPbCl3} is obtained by optimizing the experimental structure. Phase-I of \ch{MHyPbBr3} is theoretically constructed from its chloride analog, followed by its optimization. $m-m'$, and $n-n'$ are used to differentiate the two type of \textit{guest} orientations in Phase-II of both the perovskites. In Phase-I, each of the \textit{guest} orientations is symmetry-equivalent. Pink/grey-colored spheres are for lead atoms in the less/more distorted octahedral layer, green is for chlorine atoms, and dark brown is for bromine atoms in the \textit{host}. The \textit{guest} comprises of nitrogen atoms represented by blue-colored spheres, hydrogen atoms represented by cream-colored spheres, and carbon atoms shown by light brown atoms.}
	\end{figure}

	Phase-II, from the experiments~\cite{maczka2020methylhydrazinium}, and Phase-I of \ch{MHyPbBr3}, which is theoretically constructed from \ch{MHyPbCl3}, were fully optimized using PBEsol+D2 functional. The theory level chosen here was previously shown to yield good agreement with experimental parameters.~\cite{pradhi_mhpc} Phase-II optimizes in the monoclinic phase (space group, $P2_1$), same as the experimental phase, and Phase-I optimizes in the orthorhombic structure, same as its chloride analog, as shown in Figs.~\ref{fig_ch5:mhpb_mhpc_ht} and~S1. The optimized structures for Phase-I and Phase-II of chloride perovskite were used from our previous work~\cite{pradhi_mhpc}. The electronic properties (such as band dispersion and DOS) obtained for Phase II are shown in Fig.~S3. The predicted band gap of $2.54$ eV matches very well with the experimentally reported value of $2.58$ eV. 
	
\begin{figure}[tbh!]
		\centering
		\subfigure{\includegraphics[width=0.45\textwidth]{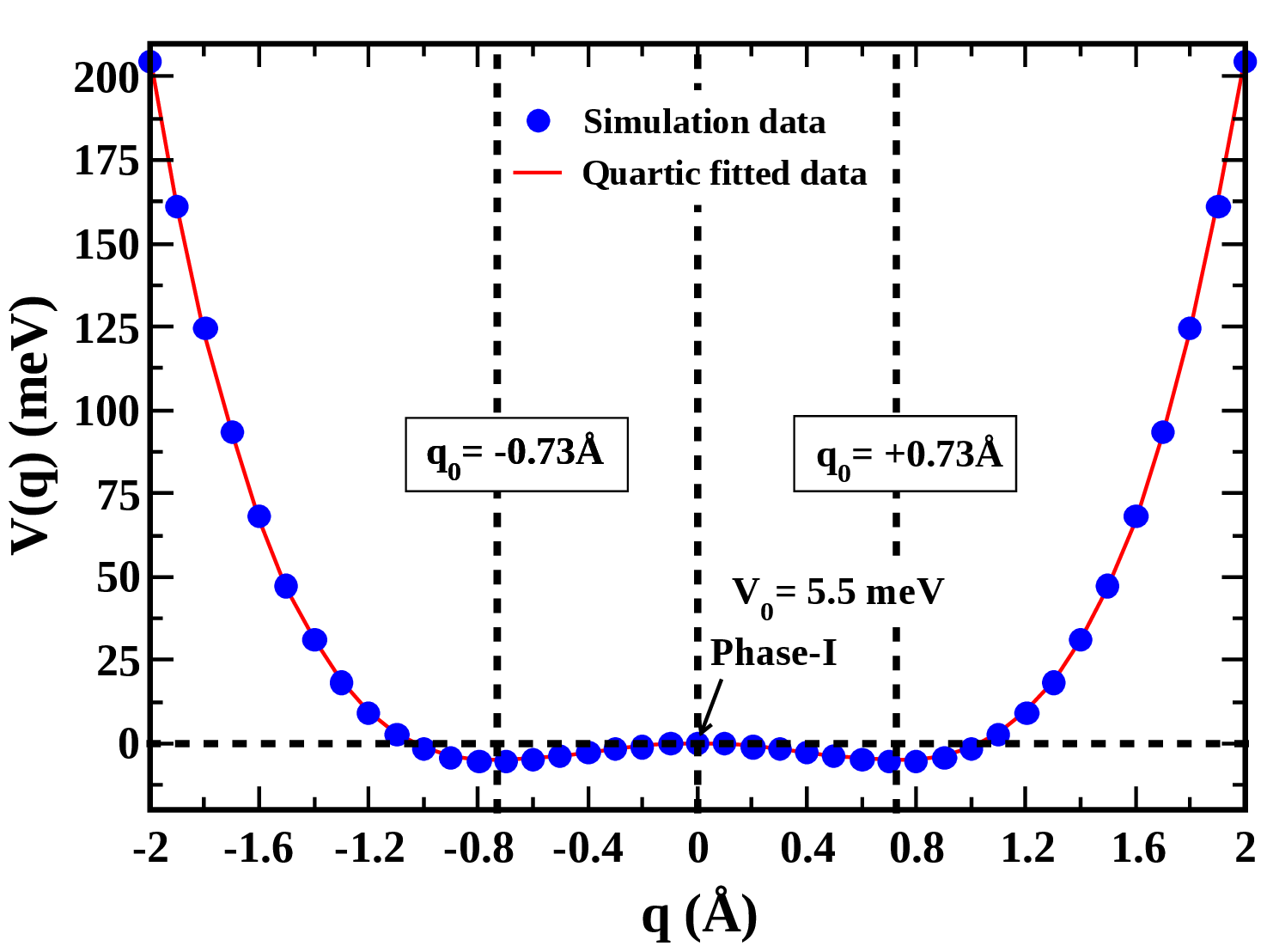}\label{subfig:mhpc_pes}}
		\hfill
		\subfigure{\includegraphics[width=0.45\textwidth]{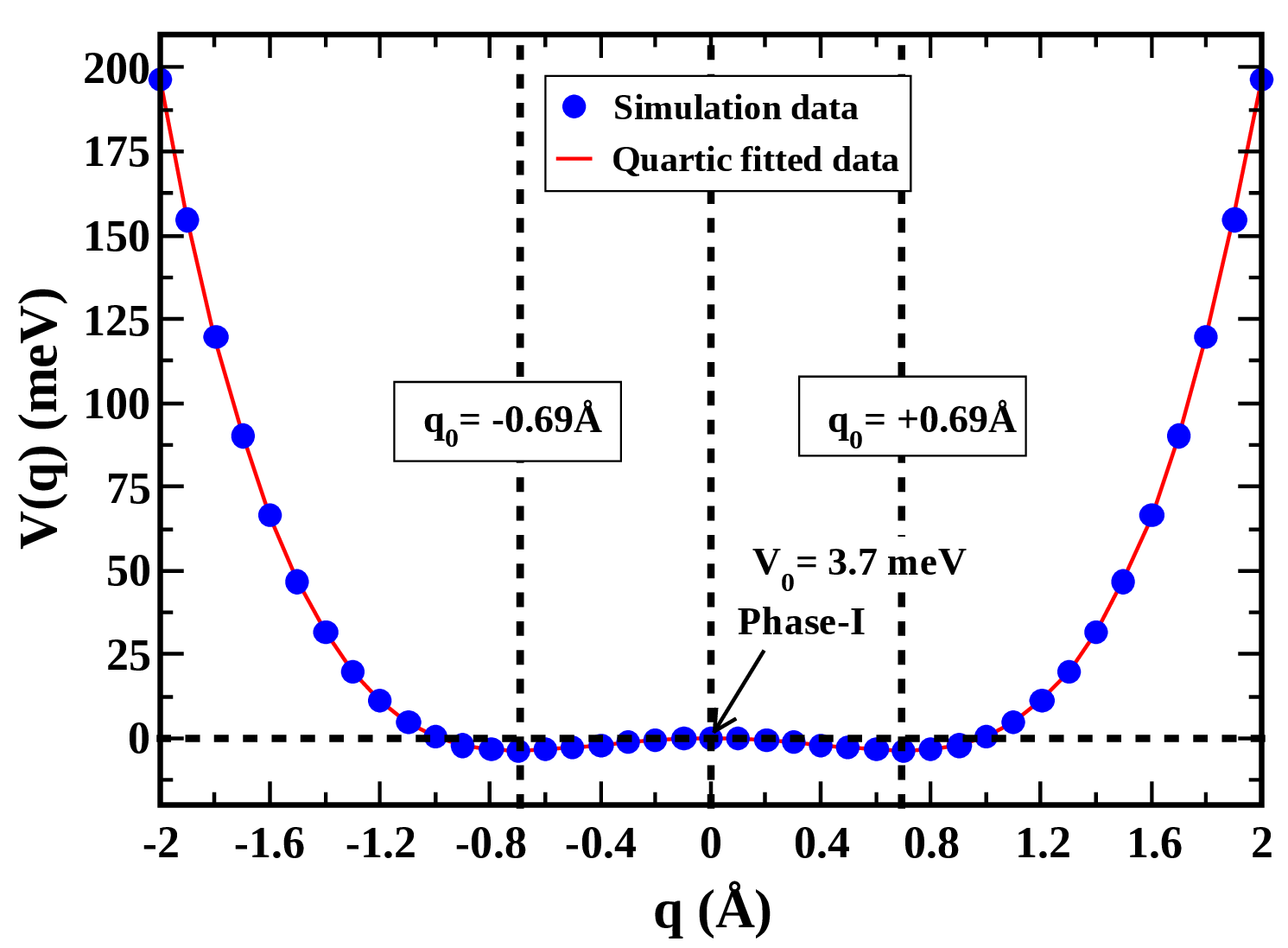}\label{subfig:mhpb_pes}}
		\caption{\label{fig_ch5:mhpb_mhpc_PES}~Potential energy curves (PEC) obtained by following the unstable mode of Phase-I of (a) \ch{MHyPbCl3} and (b) \ch{MHyPbBr3}. The blue circles on the plot refer to the single-point energy values as obtained from DFT. The red curve is the quartic fit to the PEC for the unstable mode of Phase-I. The barrier height ($V_0$) and position of the off-center minima ($\pm q_0$) are indicated in the plots.}
	\end{figure}

	A dynamical stability analysis for the optimized structures (Fig.~S2) revealed that, while Phase-II remains dynamically stable at $0$ K, the optimized Phase-I of \ch{MHyPbBr3} is dynamically unstable with an imaginary frequency of $0.6$ THz, similar to the instability of Phase-I of \ch{MHyPbCl3}.~\cite{pradhi_mhpc} This instability is resolved by following the unstable mode for both the halide perovskites, as inferred from the potential energy curve (PEC) along this mode shown in Fig.~\ref{fig_ch5:mhpb_mhpc_PES}. The structures optimized around the minima of the PEC are found to be dynamically stable. Two energy-equivalent minima along this PEC are separated by a barrier of $\sim$5 meV or lower. This indicates that the experimentally observed structure of Phase-I is likely a result of dynamical disorder over the two oppositely displaced structures. This noteworthy feature revealed by the present lattice dynamical analysis has not been experimentally anticipated to the best of our knowledge. 
	
	\subsection{Qualitative model for transition in \textit{host/guest} systems}
	
\begin{figure}[tbh!]
		\centering
		\includegraphics[width=0.4\textwidth]{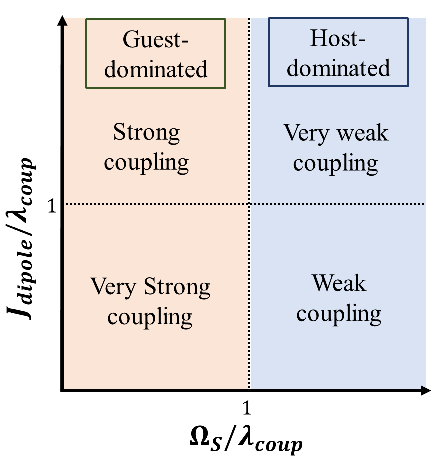}
		\caption{\label{fig:coup_phase_dig}~Phase diagram for different regimes of coupling. The orange-shaded region exhibits strong \textit{guest/host} coupling, making \textit{guest}, and the transitions in these regions are dominated by \textit{guest}-reorientation. In other words, \textit{guest} is the primary order parameter of structural phase transition. The blue-shaded regions exhibit weak HG coupling, thereby making \textit{host} the driving force for structural phase transition.}
	\end{figure}

	The sequence of events taking place during temperature-induced phase transition can be qualitatively explained on the basis of competition among three parameters: $\Omega_{S}$ indicating the strength of \textit{host} elasticity or the depth of the potential well separating the distorted and the undistorted phases, the \textit{HG} coupling strength denoted by $\lambda_{coup}$, and dipole-dipole coupling strength between the neighboring \textit{guests} denoted by $J_{dipole}$. Different regimes arise depending on the relative strengths of these interactions, where either sub-lattice drives transitions, as is shown in Fig.~\ref{fig:coup_phase_dig}. The energies associated with these constants can be estimated using our DFT methods to assess their relative strengths. The definitions of elastic energy cost ($E_{elas}^A$), associated with $\Omega_S$, and the \textit{host/guest} interaction energy ($E_{HG}^{ind}$) associated with $\lambda_{coup}$, were detailed in our previous work~\cite{pradhi_mhpc} (also see Fig.~S6 and Table~S3).
	
	We compare the relative magnitudes of $E^{ind}_{HG}$ and $E^{A}_{elas}$ for both the halide perovskites, assuming that the dipole-dipole coupling strength will be similar in the two cases. The magnitude of $E^{ind}_{HG}$ across a range of $A-$site ions and across halide ions (Br and Cl) for \ch{MHy+} is tabulated in Table~S4, which shows a high HG coupling strengths for MHy-containing perovskites. We expect that if $\lambda_{coup}>\Omega_S$ (or alternatively, $E^{ind}_{HG}>E^{A}_{elas}$), then the \textit{host} should first transition to an undistorted state, followed by the disordering of the \textit{guest}. On the other hand, if $\lambda_{coup}\approx \Omega_{S}$, then the \textit{host} distortion and \textit{HG} coupling will vanish simultaneously. Without the latter coupling, \textbf{}the \textit{guest} disordering would depend entirely on $J_{dipole}$ which is typically lower than $\lambda_{coup}$ for these hybrid perovskites (see Sec.~SI-C). Hence, the resulting sequence of phase transition would be from an ordered state directly to the disordered cubic phase. The ordered phase would have a wider (narrower) stability range depending on how large (small) $E^{ind}_{HG}/E^{A}_{elas}$ is compared to unity.
	
	\begin{table}
		\caption{The \textit{host/guest} interaction energy and ($E^{ind}_{HG}$), \textit{host} elastic energy cost ($E^{A}_{elas}$) and their ratio for chloride and bromide perovskites for Phase-I \textit{guest} orientation.}
		\label{tab:hg_inter_strength_phaseI}
		\centering
		\begin{tabular}{l|cc}
			\hline
			\hline
			&\ch{MHyPbCl3}&\ch{MHyPbBr3}\\
			\hline
			\hline
			$E^{ind}_{HG}$ (eV/f.u.) &2.83&2.99\\
			$E^{A}_{elas}$ (eV/f.u.) &2.52&2.85\\
			$E^{ind}_{HG} / E^{A}_{elas} $&1.13&1.05\\
			
			\hline
			\hline
		\end{tabular}
	\end{table}
	
	We look at the the transition of the two ordered phases (stable states for Phase-I \textit{guest} configuration, and Phase-II \textit{guest} configuration), separately. Table~\ref{tab:hg_inter_strength_phaseI} shows the ratio of the two competing interactions, $E^{ind}_{HG}/E^{A}_{elas}$, for the first case. The stable equilibrium state, in this case, is the dynamically stable Phase-I, obtained after following the unstable phonon mode. A ratio of greater than 1 for chloride implies that Phase-I gets stabilized over the cubic phase because of a relatively stronger HG interaction than the \textit{host} elasticity. On the other hand, this ratio is nearly one for bromide perovskite, implying that the ordered Phase-I has no extra stabilization over the cubic phase. Hence, the completely disordered state is preferred in bromide, thereby explaining the absence of Phase-I from its phase diagram. For the transition from Phase-II to cubic, the corresponding ratio of coupling constants (see Table~S5) shows a larger deviation from unity as compared to that for Phase-I \textit{guest} configuration of both halides. This implies that Phase-II has a larger stability range than Phase-I in chloride and the disordered phase in bromide perovskite. This makes Phase-II the most stable structure, hence the ground state in both the perovskites, in agreement with the experimental observations~\cite{maczka2020methylhydrazinium,maczka2020three}. A more quantitative explanation is detailed in the sections below.
	
	\subsection{Phase-I in \ch{MHyPbX3} (\ch{X=Cl,~Br})}
	
	Despite being structurally similar, \ch{MHyPbCl3} and \ch{MHyPbBr3} display different phases at higher temperatures, as has been demonstrated experimentally~\cite{maczka2020methylhydrazinium,maczka2020three}. The high-temperature, more polar phase, Phase-I, is absent in bromide perovskite, whereas it occurs at around $340$ K in chloride analog. To understand the origin of this difference, we computed the relative free energies of two phases, Phase-I and Phase-II, in \ch{MHyPbX3} (\ch{X=Cl,~Br}) as a function of temperature. For this, we determined the thermal properties from the phonon frequencies for both the compounds of \ch{MHyPbX3}. To account for thermal fluctuations along the unstable eigenmode of Phase-I we implemented a correction to the free energy, as explained below, which helps estimate a fairly accurate phase transition temperature (from Phase-II to Phase-I).
	
	The phase transition temperature can be obtained by comparing the Gibbs free energy, $\mathcal{G}$, of the two phases as a function of temperature~\cite{saidi2016nature,lohaus2020thermodynamic,zhang2018intrinsic}. In general, the Gibbs free energy is defined as $\mathcal{G}=U+pV-TS$, where $U$ is the internal energy, $p$ is the pressure experienced by the system of volume $V$, $T$ is the temperature, and $S$ is the entropy of the system. $U$ includes both the electronic energy and the zero-point energy of the phonons. 
	For the completely optimized state, $p\to0$, and hence $\mathcal{G}=\mathcal{F}=U-TS$, where $\mathcal{F}$ is Helmholtz free energy (free energy at constant volume). The free energy can also be defined in terms of partition function as $\mathcal{F}=-k_BT~\text{ln} (Z)$, $Z$ being the partition function comprising of all the states (3$N$ phonon modes, where $N$ is the number of atoms in the system)~\cite{baroni2001phonons}. The application of a harmonic model to estimate free energies requires all the phonon modes to be stable. We have an unstable mode in the phonon dispersion of Phase-I, which would usually just be dropped out from the partition function. However, given that the unstable mode is responsible for dynamical disorder in the crystal, we propose to account for it using the DFT potential energy curves. Since the difference in the volume of Phase-II (low temperature) and Phase-I (high temperature) structures is negligible (less than $2\%$), we can ignore any volume-dependence of phonon frequencies. Then, the modified $\mathcal{G}$ can be expressed as follows:
	
	\begin{equation}\label{eq_ch4:mod_gibbs_en}
		\mathcal{G}=-k_BT\sum_{n=1}^{3N-1} \text{ln}(Z_n)-k_BT~\text{ln}(Z_{unstable})
	\end{equation}
	
	As can be seen from Fig.~\ref{fig_ch5:mhpb_mhpc_PES}, the PEC of the unstable mode of Phase-I in both the perovskites is of the form of a double well potential, which can be modeled by a quartic form $V(q)=a_1q^2+a_2q^4$, with $a_1<0$ and $a_2>0$. The fitted data is shown by the red curve in Fig.~\ref{fig_ch5:mhpb_mhpc_PES}. The values of the parameters, $a_1$ and $a_2$, thus obtained have been tabulated in Table~S2. The classical partition function, in general, is given by $Z_{unstable}=2\int_{0}^{\infty}e^{-\beta V(q)}$, the factor of 2 is due to the symmetry of the quartic potential about $q=0$. The exact exponential for the quartic form of potential is integrated numerically to get the partition function corresponding to the unstable mode. In what follows, we refer to the inclusion of this term as the {\it quartic correction}.
	
\begin{figure}[tbh!]
		\centering
		\subfigure[\ch{MHyPbCl3}]{\includegraphics[width=0.4\textwidth]{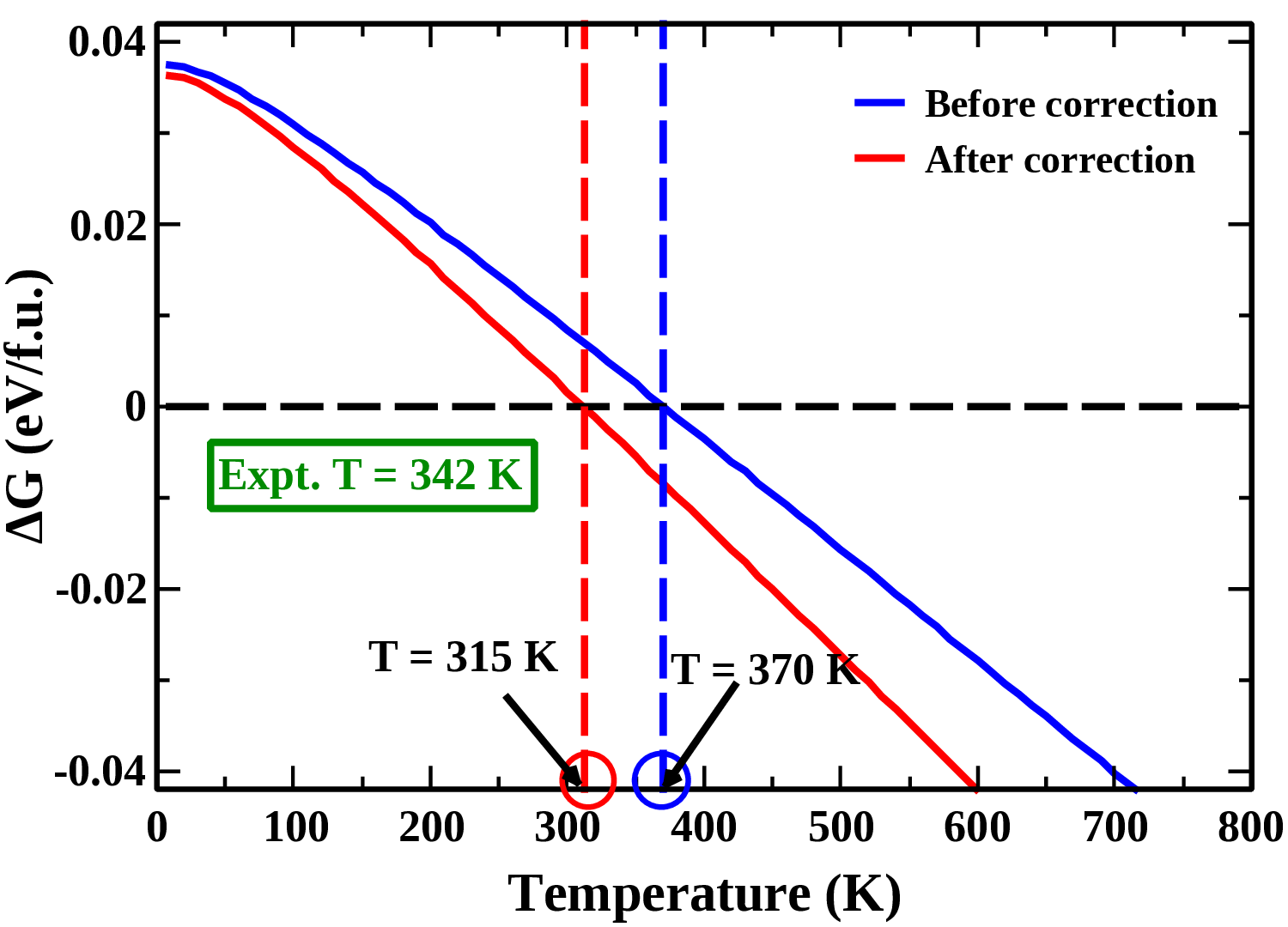}\label{subfig:fen_corrected_mhpc}}
		\hfill
		\subfigure[\ch{MHyPbBr3}]{\includegraphics[width=0.4\textwidth]{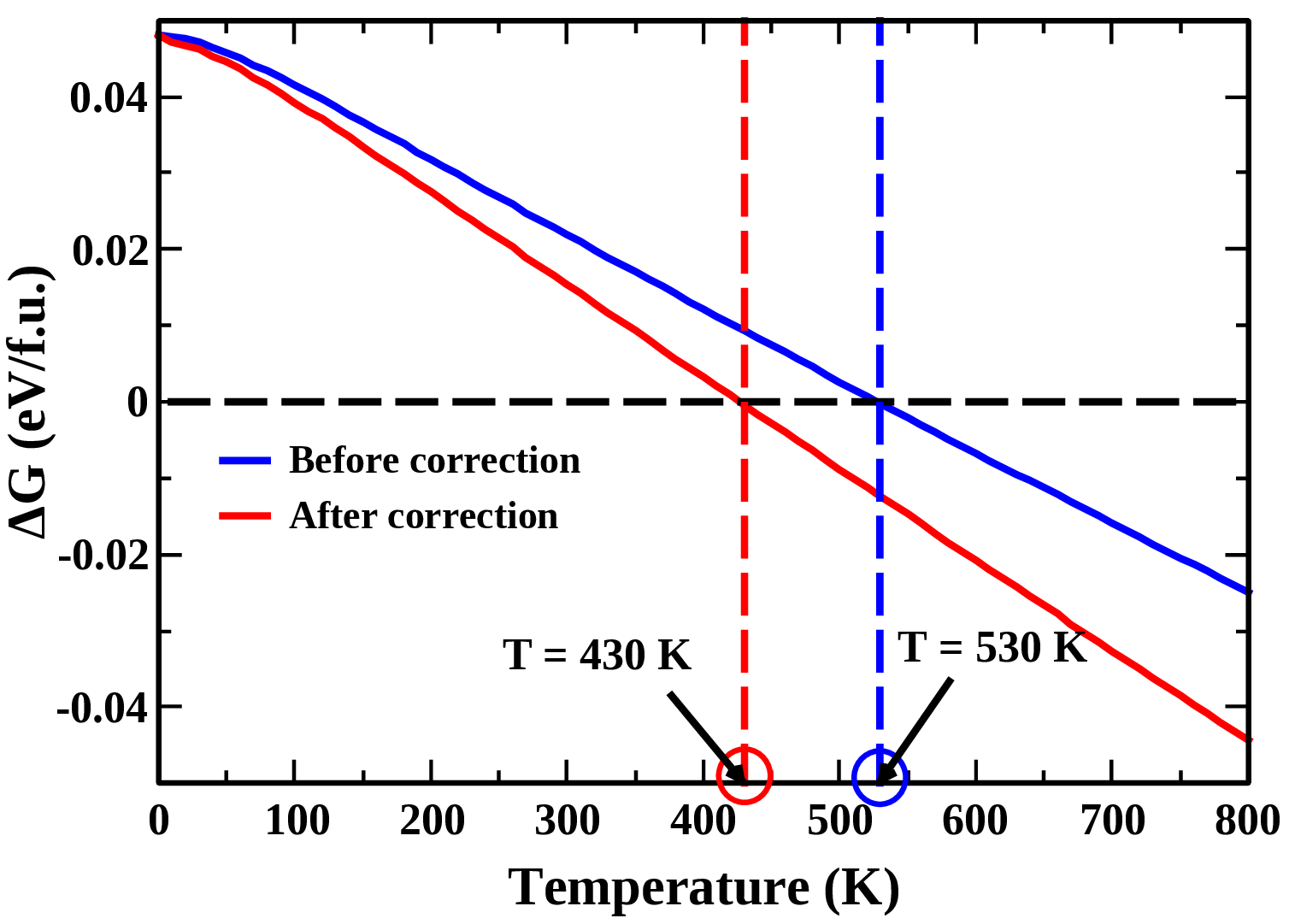}\label{subfig:fen_corrected_mhpb}}
		\caption{\label{fig_ch4:fen_corrected_mhpc_mhpb}~Corrected $\Delta G$ for the phase transition temperature between the Phase-II and Phase-I of \ch{MHyPbX3}.}
	\end{figure}

	The variation of $\mathcal{G}$ of the two polar phases defined by $\Delta \mathcal{G}=\mathcal{G}_{Phase-I}-\mathcal{G}_{Phase-II}$, is plotted as a function of temperature before and after incorporating the quartic correction for both the perovskites in Fig.~\ref{fig_ch4:fen_corrected_mhpc_mhpb}. $\Delta \mathcal{G}$ goes to zero at the phase transition temperature. In both halides, we find that the quartic correction lowers the transition temperature from Phase-II to Phase-I, as expected by the increased entropy associated with the dynamical disorder. We find that the transition in the case of chloride takes place around $315$ K (after introducing the correction), which is very close to the experimental value of $342$ K, as is shown in Fig.~\ref{subfig:fen_corrected_mhpc}. Surprisingly, the quartic corrected transition temperature corresponding to the Phase-I of \ch{MHyPbBr3} was found to be $430$ K as shown in Fig.~\ref{subfig:fen_corrected_mhpb}, quite close to the cubic transition temperature of $420$ K. In analogy with the chloride perovskite, this temperature could be an underestimate, implying that the cubic phase becomes more stable than Phase-I of bromide perovskite before $430$ K. Hence, the latter is not experimentally observed. However, the comparison of free energies with the cubic phase by a similar method is not possible because of the presence of several dynamical instabilities in the $0$ K cubic structure (and since the actual phase is disordered). Alternatively, the cubic phase in bromide perovskite could derive from disordering over orthorhombic distortions represented locally by the Phase-I structure. The free energy of the cubic phase in a similar fashion as outlined above was not possible given the numerous unstable modes occurring in the optimized cubic structure as well as the expected disorder. We analyzed the latter possibility through AIMD simulations, as detailed in the next section.

	Interestingly, the dynamic instability in Phase-I vanishes at higher pressures, removing the disordered nature of the phase. A plot of the square of the soft mode frequency with pressure (see Fig.~S15) reveals that the mode stabilizes at $\sim$4 GPa in \ch{MHyPbCl3} and at $\sim$5 GPa for Br analog (see also Figs.~S12 and S13). Hence, we anticipate that the new phase that emerges at higher pressure would find the dynamic disorder seen in Phase-I suppressed.

	\subsection{Cubic phase of \ch{MHyPbX3}}
	
	In order to probe the nature of the cubic phase of \ch{MHyPbBr3}, we carried out AIMD simulations on the cubic phase under the constraints of experimental volume and temperature of 470~K. These simulations are primarily aimed at establishing the thermodynamically favored MHy orientations that are sampled in the cubic phase. Such an identification allowed us to not only rationalize the stability of the cubic phase in the bromide perovskite, but also to rule it out in the chloride analog. Furthermore, being a larger \textit{guest}, \ch{MHy} is expected to demonstrate slower reorientation dynamics in comparison to other known organic cations commonly used in OIHPs.~\cite{pradhi_mhpc} The role of the hydrogen and coordinate bonding of \textit{host} and \textit{guest} on this slowdown can also be investigated through these simulations.
	
	We created a 2x2x2 supercell of the cubic unit cell with the experimental lattice parameters by choosing random starting configurations for each of the \textit{guest}s (see Fig.~S16). An alternative starting point yielded a similar sampling of orientational phase space (see Fig.~S33), indicating the insensitivity of the sampling to the initial \textit{guest} orientations. In addition to simulations where both \textit{host} and \textit{guest} were allowed to move, we also performed simulations in which the \textit{host} was constrained to be in the ideal cubic geometry. We term these as flexible and frozen \textit{host} simulations, respectively, in what follows.
	
	\subsubsection{\textit{Guest} orientation parameters}
	\begin{figure*}
		\centering
		\subfigure[]{\includegraphics[width=0.45\textwidth]{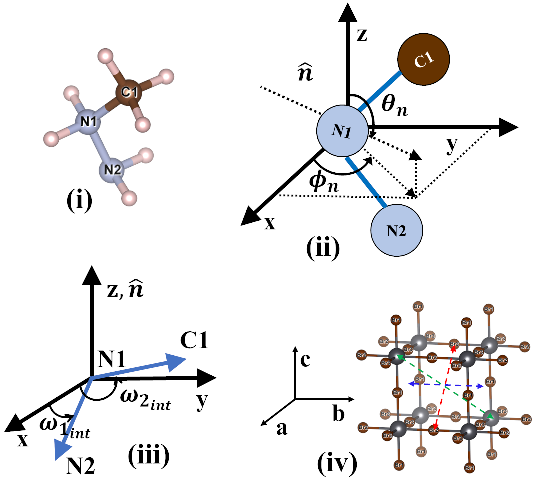}\label{subfig_ch4:schematic_guest}}
		\hfill
		\subfigure[]{\includegraphics[width=0.45\textwidth]{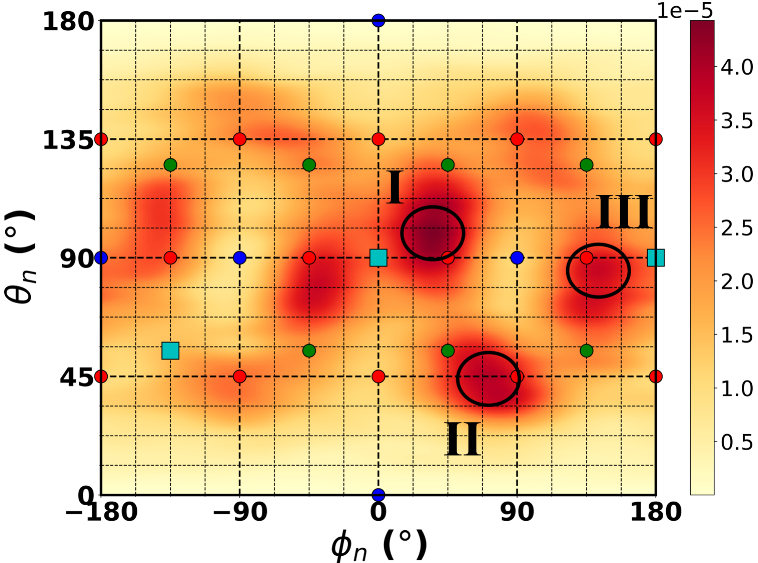}\label{subfig_ch4:flex1_norm_orient}}
		\hfill
		\hfill
		\subfigure[]{\includegraphics[width=0.45\textwidth]{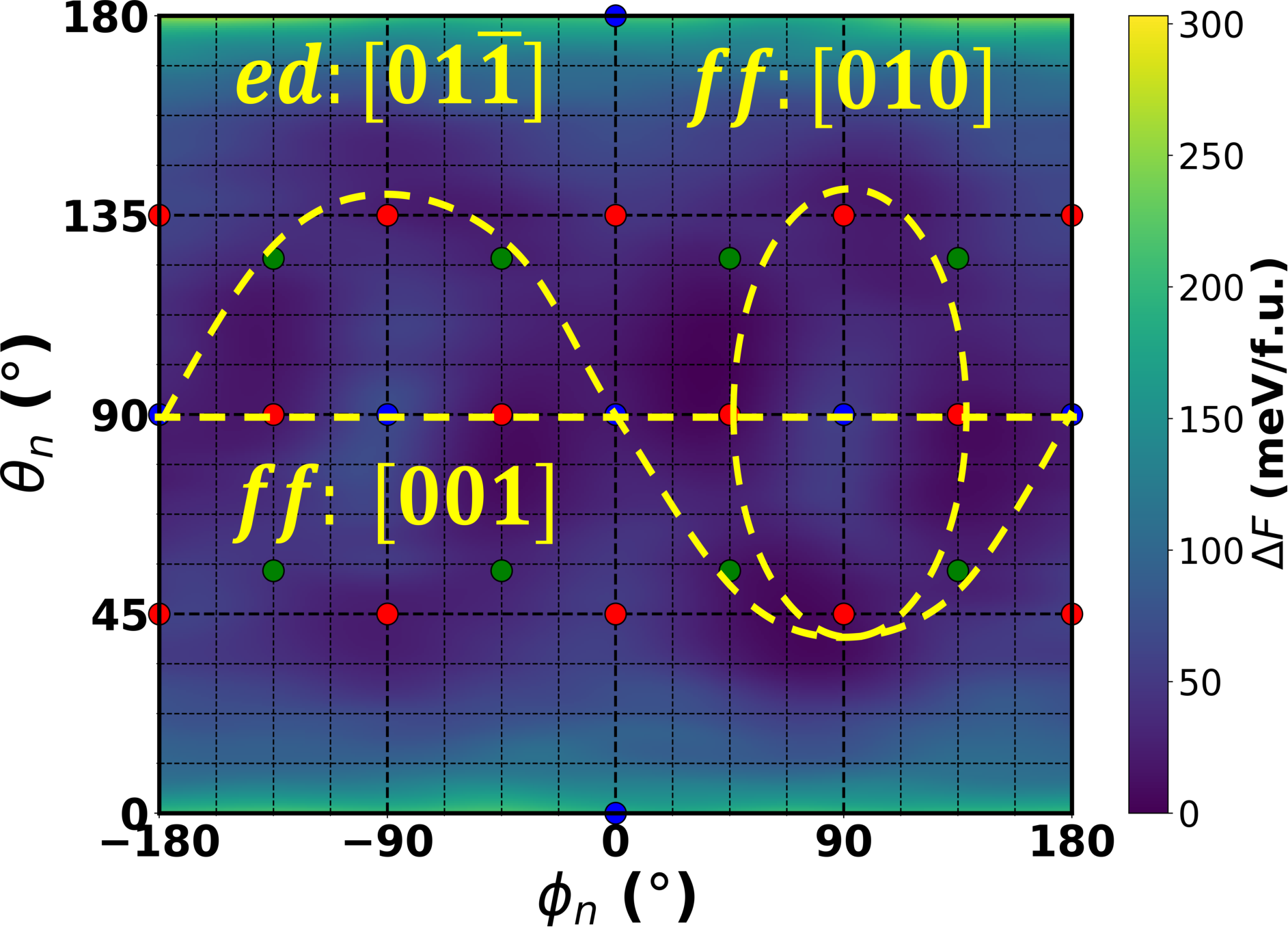}\label{subfig_ch4:flex1_free_en}}
		\hfill
		\subfigure[]{\includegraphics[width=0.45\textwidth]{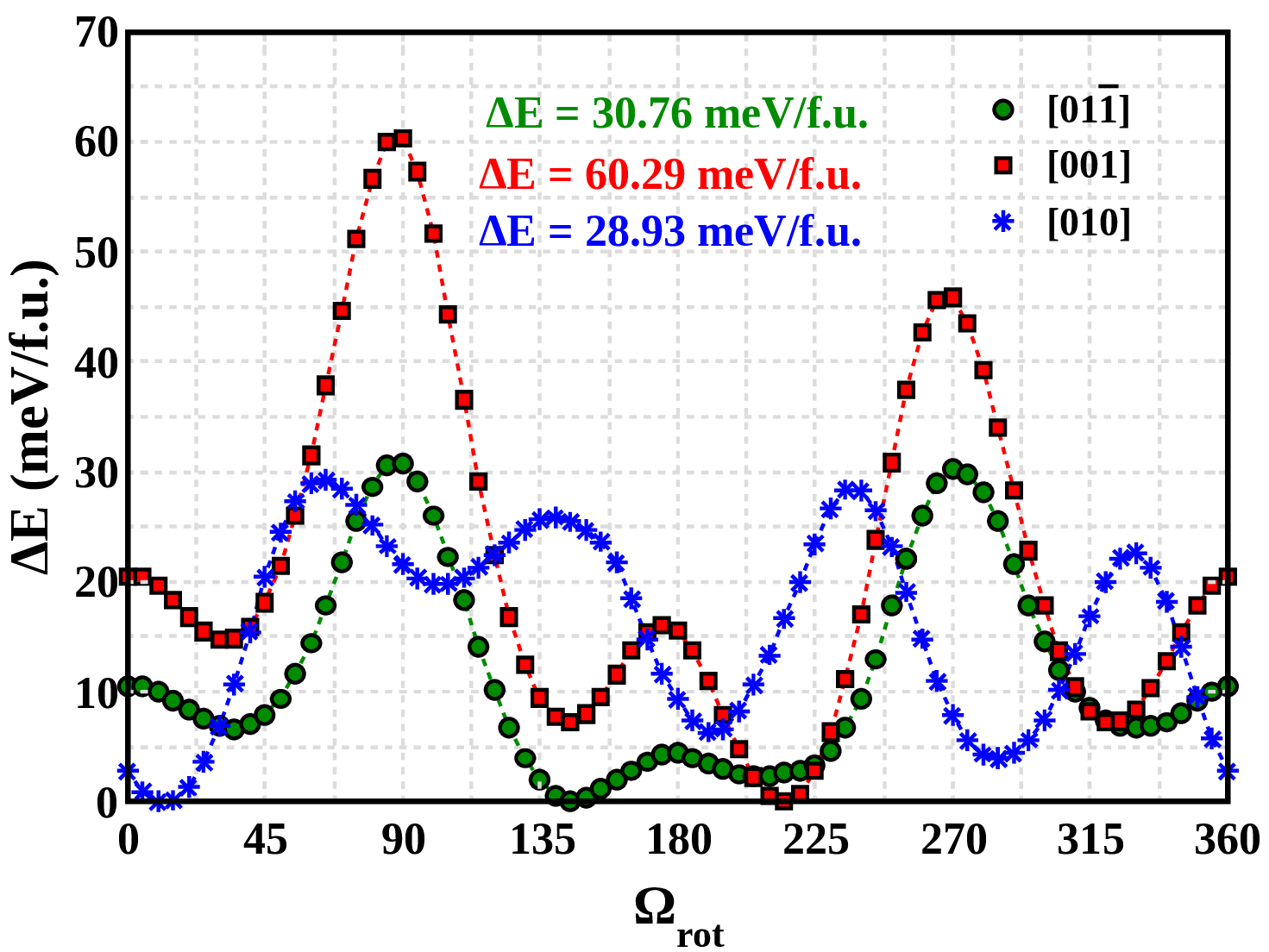}\label{subfig_ch4:flex1_free_en_2D}}
		\caption{\label{fig_ch4:flex1_or_free_en}~\subref{subfig_ch4:schematic_guest} A description of the coordinate system employed: (i) the \textit{guest} molecule, \ch{MHy+}, (ii) the instantaneous orientation of the normal to its molecular plane ($\hat{n}$), (iii) the orientation of its bonds with respect to a fixed reference axes in the molecular plane. Here the normal is oriented along the $z$-axis, such that the molecule lies in the $xy$-plane, and (iv) high symmetry directions in the cubic phase indicated by blue (face to face (\textit{ff})), green (body diagonal (\textit{bd})) and red (edge diagonal (\textit{ed})); ~\subref{subfig_ch4:flex1_norm_orient}~ orientational distribution of the molecular plane-normals, $\hat{n}~(\phi_n,~\theta_n)$; ~\subref{subfig_ch4:flex1_free_en}~distribution of the orientational free energy (meV/f.u.) associated with \ch{MHy} orientations at $470$ K. Yellow dashed curves indicate rotation pathways about three symmetry axes ($ed:[01\bar{1}],~ff:[001]$, and $ff:[011]$) along which energy equivalent \textit{guest} orientations can be achieved; \subref{subfig_ch4:flex1_free_en_2D}~free energy profile for rotation about the $[0\overline{1}1]$ (green curve), the $[001]$ (red curve), and the $[010]$ (blue curve) axes. Blue, green, and red circles in the distribution plots correspond to \textit{ff}, \textit{bd}, and \textit{ed} orientations. Cyan-colored squares represent the starting orientations of normal to the molecular plane.}
	\end{figure*}
	
	The instantaneous orientation of any planar MHy$^+$ ion molecule is specified, as shown in Fig.~\ref{subfig_ch4:schematic_guest}, by the plane-normal $\mathbf{\hat{n}}$ and the angles, $\omega_{int}$, which represent rotations of the N1-N2 and N1-C1 bonds with respect to a reference axis in the molecular plane (also see Sec.~S-II A).
	The probability distribution function (PDF) of the molecular plane orientations over the entire trajectory (Fig.~\ref{subfig_ch4:flex1_norm_orient}) reveals that the \textit{guest} is disordered over a finite number of preferred orientations in the cubic phase. The high symmetry directions of the cube (Fig.~\ref{subfig_ch4:schematic_guest} (iv)), represented by blue, green, and red-filled circles, are also indicated for comparison. 
	The free energy landscape associated with the MHy orientations, shown in Fig.~\ref{subfig_ch4:flex1_free_en},
	indicates that the preferred orientations are separated by barriers of 40-50 meV/f.u. This is also confirmed by computing the free energies along a path of rotation about various symmetry axes (see SM for details), as shown in Fig.~\ref{subfig_ch4:flex1_free_en_2D}. Given the thermal energy available at $470$ K, it should then be highly probable for the cations to switch between different preferential sites, which implies a disordering of the \textit{guests}. Further evidence of the dynamic nature of the disorder is obtained from the rotational auto-correlation function associated with $\mathbf{\hat{n}}$ and has been discussed in the next section. 
	PDF of other internal parameters, such as the bond angle ($\angle C1N1N2$), bond lengths ($d_{C1N1}$ and $d_{N1N2}$), the shift in the center of mass ($\Delta COM$) with respect to the center of the cube $(0.5, 0.5, 0.5)$, are shown in Figs.~S17--S19. Kernel Density Estimate (KDE) was used to smoothen all PDFs appearing in this work unless specified otherwise. The average values for the internal parameters have been tabulated in Table~S8. 
	
	\subsubsection{Coexistence of cubic and orthorhombic phases}
\begin{figure}[tbh!]
		\centering
		\includegraphics[width=\columnwidth]{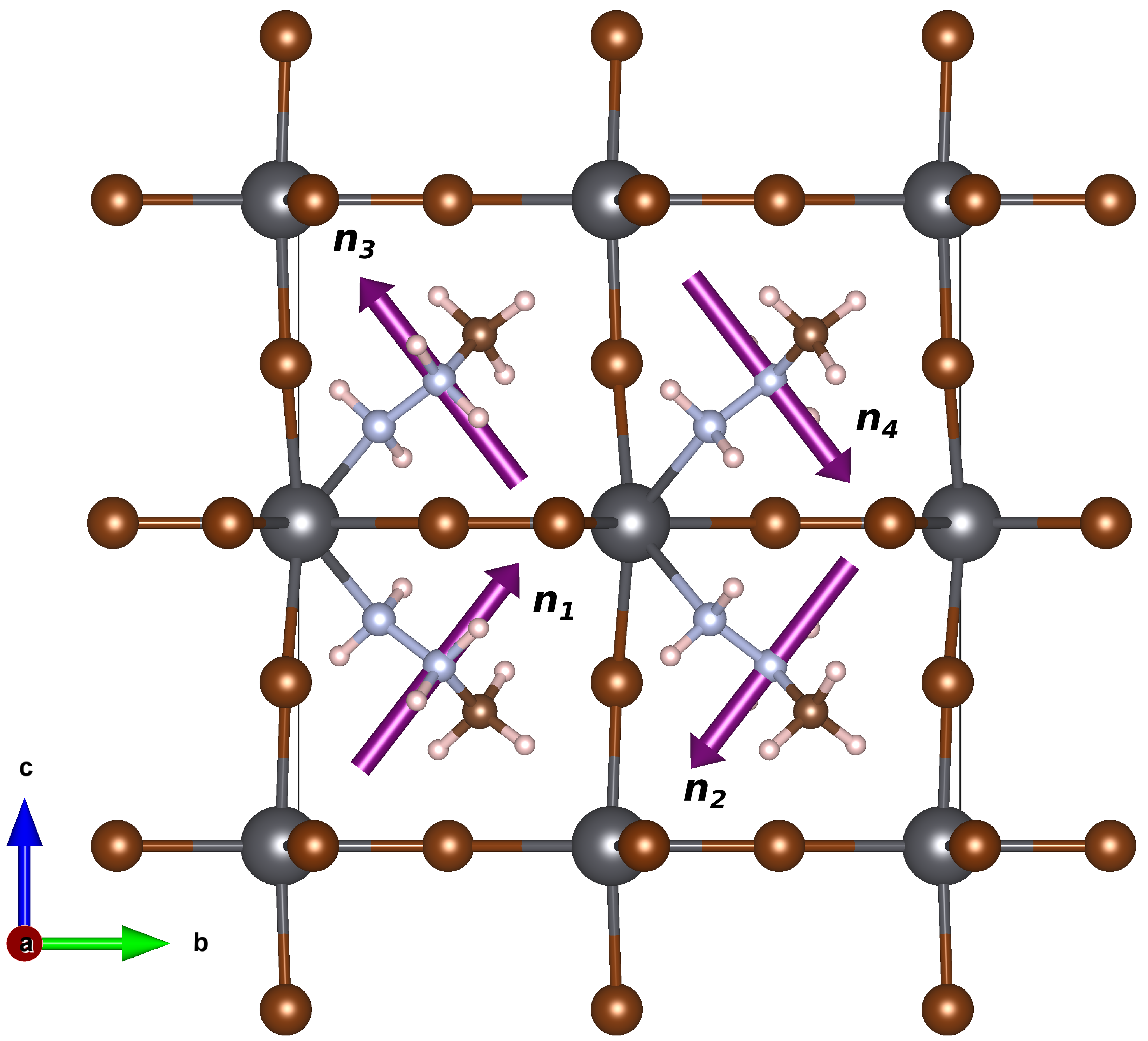}
		\caption{Violet arrows represent the normal vectors to each of the \textit{guest} ions in the orthorhombic phase of \ch{MHyPbBr3}. The labeling of the cells and the \textit{guest} ions is according to the numbering of the normal vectors, $\mathbf{n}$. $\mathbf{n_1}$ sits at the bottom left corner (cell number, \textbf{l}=(0,0)), $\mathbf{n_2}$ is bottom right corner corresponding to \textbf{l}=(1,0), $\mathbf{n_3}$ sits st the top left corner corresponding to \textbf{l}=(0,1), and $\mathbf{n_4}$ is at the top right corner corresponding to \textbf{l}=(1,1) in the $bc$-plane.}
		\label{fig:norm_to_ortho_phase}
	\end{figure}

	The orthorhombic (Phase-I) unit cell comprises of four units of \ch{MHyPbBr3} in the $bc$-plane with the plane-normals (denoted by {\bf n}$_i$) of \textit{guest} ions oriented as shown in Fig.~\ref{fig:norm_to_ortho_phase}. We note here that the angle between the plane-normals of two adjacent MHy along $c$ axis (e.g. {\bf n}$_1$ and {\bf n}$_3$) in the optimized structure was $\phi = 0.46\pi$. This pattern is repeated along the \textit{a}-axis. All \ch{Pb...N} coordinate bonds are of the same length ($2.78~\text{\AA}$), a feature that is also retained in the cubic phase simulated here (see Fig.~S24). We define below a layerwise order parameter (OP), $\xi_{a,i}$, to keep track of the relative ordering of the neighboring \textit{guest} ions in the $i^{\rm th}$ layer.
	\begin{align}
		\xi_{a,i} &=  {\overrightarrow{\eta}}_{\textbf{l}_1}^a \cdot \overrightarrow{\Omega}^{a,i} \\
		\overrightarrow{\Omega}^{a,i} &=\frac{1}{4}\sum\limits_{k=1}^4 U(\theta_{{\textbf l}_k})\cdot{\overrightarrow{\eta}}^a_{{\textbf l}_k} 
	\end{align}
	where ${\overrightarrow{\eta}}_{\textbf{l}_k}^a$ is the projection of {\bf n}$_k$ on the $bc$-plane, ${\textbf l}_k$ are lattice vectors in the $i^{\rm th}$ layer representing the four unit cells shown in Fig.~\ref{fig:norm_to_ortho_phase} and $\theta_{{\textbf l}_k}= \vec{\kappa} \cdot {\textbf l}_k$. The wave-vector $\vec{\kappa} = [\pi,\pm\phi]$ is used depending on whether the instantaneous plane-normal in the home-cell (k=1) resembles ${\bf n}_1$ or ${\bf n}_3$. The rotation matrix, $U(\theta_{\textbf{l}_k})$, can be defined as:	
	\[
	U(\theta_{\textbf{l}_k})	= \begin{pmatrix}
		cos(\theta_{\textbf{l}_k}) &-sin(\theta_{\textbf{l}_k})\\
		sin(\theta_{\textbf{l}_k}) & cos(\theta_{\textbf{l}_k})\\
	\end{pmatrix}
	\]
	
	The layerwise OP, $\xi_{a,i}$, reports on the similarity of the collective orientation of \textit{guest} ions in a layer to the reference (orthorhombic phase) orientation. Large absolute values indicate similarity to the ordering in the orthorhombic phase. Ordering across layers can be tracked by defining the global OP
	\begin{align}
		\xi_a^\pm = \left(\xi_{a,1}\pm\xi_{a,2}\right)/2
	\end{align} 
	where $\xi_a^\pm \in \left[-1,1\right]$ measures the in-phase ($+$) or out-of-phase ($-$) cooperative alignment of the \textit{guest} ions across the two layers along $a$ axis in the supercell considered. Analogous order parameters can be defined for the $ca$ ($\xi_b$) and the $ab$-planes ($\xi_c$). A more general method for such order parameters was outlined in a previous work.~\cite{maity2024cooperative} A large value of $\left\vert{\xi_a^+}\right\vert$ (close to one) would indicate that not only do the \textit{guest} ions in a layer resemble the Phase-I ordering, but they are also perfectly correlated along the layers as is expected in that phase. In contrast, the value of $\xi_a^-$ should go to zero for such a scenario. The value $\xi=-1$ refers to an ordered arrangement, referred to as {\bf n}$_3$-like below, which can be obtained by simply translating the arrangement in Fig.~\ref{fig:norm_to_ortho_phase} by half a unit cell along $c$-axis and, hence, is equivalent to Phase-I. 
	
\begin{figure}[tbh!]
		\centering
		\includegraphics[width=\columnwidth]{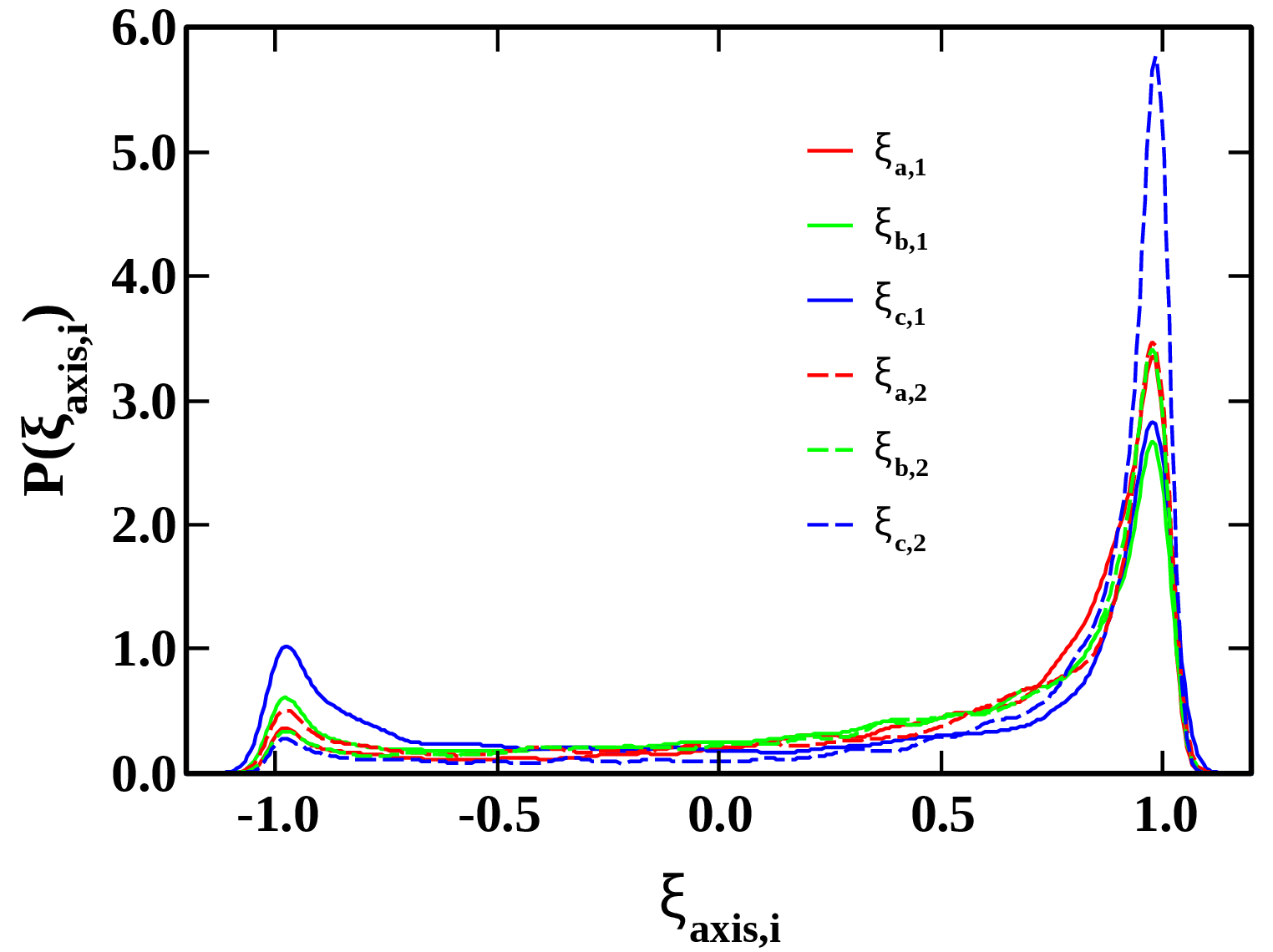}
		\caption{Probability distribution function of the layer-wise order parameter, $\xi$, along each axis. The solid curves correspond to the first layer along the axis, and the dotted curve is for the second layer. Red, green, and blue-colored lines are used to plot order-parameter values along $a,~b$, and $c$-axes, respectively.}
		\label{fig:xi_layerwise}
	\end{figure}

	Fig.~\ref{fig:xi_layerwise} shows the layerwise distribution of the order parameter in the cubic phase along all three axes. A peak around 1 indicates an orthorhombic phase-like ordering of the \textit{organic} guests in the layers (e.g. $ab$-plane for $\xi_c = 1$). During the course of the MD, the system visits orthorhombic structures of all 3 orientations, albeit unequally. The dissimilar distribution of normals over the three axes (also evident from the time correlation plot of the layerwise order parameter in Fig.~S25), could be a result of smaller size of the supercell as well as the volume constraint employed. We expect that cell flexibility would allow for the \textit{guest} molecules to equally span all three directions. In fact, we also see {\bf n}$_3$-like layers forming infrequently, which are quite likely higher energy local fluctuations from the orthorhombic global arrangement (see below). 
	
\begin{figure}[tbh!]
		\centering
		\includegraphics[width=\columnwidth]{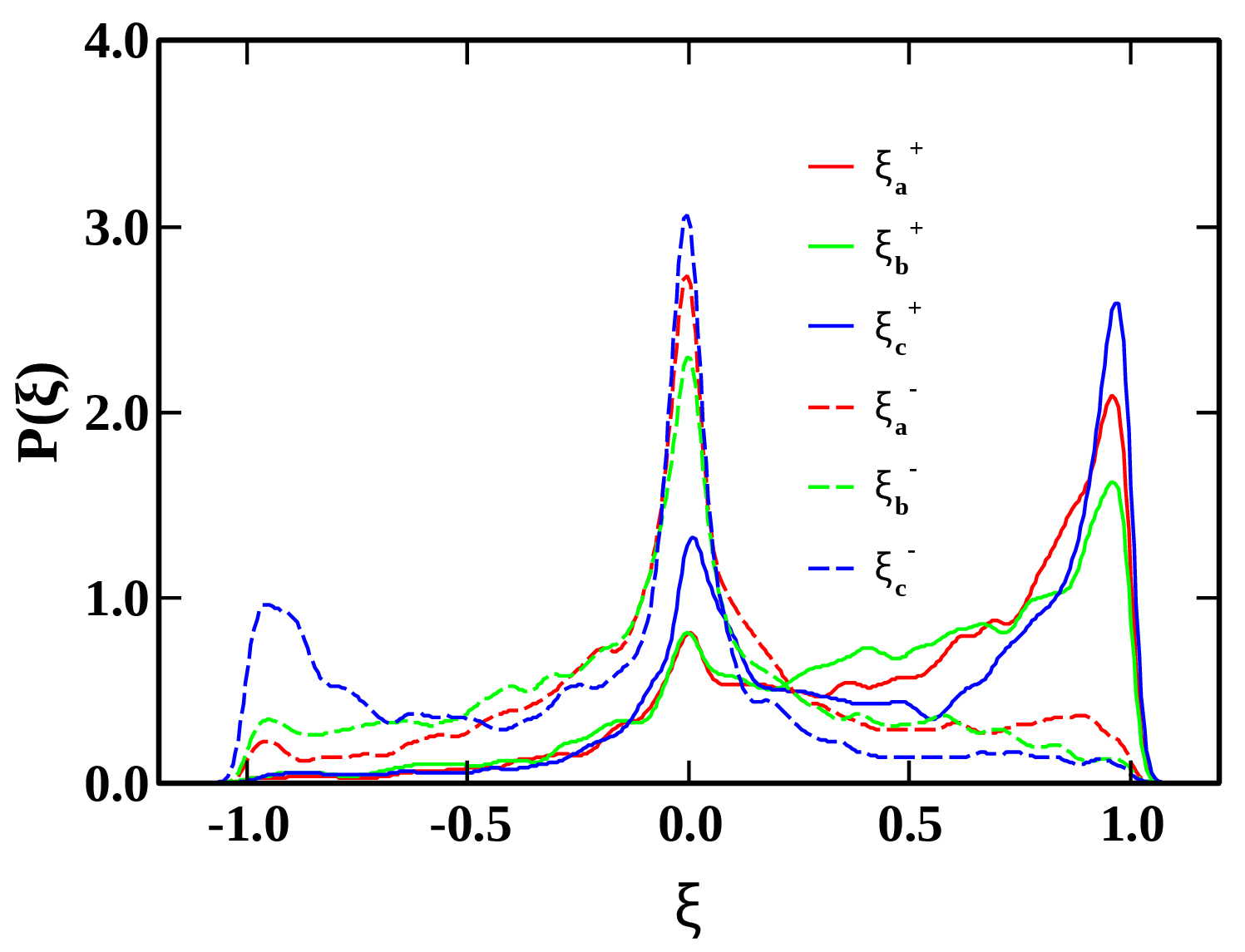}
		\caption{Probability distribution function of the in-phase and out-of-phase order parameter, $\xi^{+/-}$, along each axis. The solid curves correspond to the in-phase order parameter (OP), $\xi^+$, along the axis, and the dotted curve is for the out-of-phase OP, $\xi^-$. Red, green, and blue-colored lines are used to plot order-parameter values along $a,~b$, and $c$-axes, respectively.}
		\label{fig:xi_plus_minus}
	\end{figure}

	Fig.~\ref{fig:xi_plus_minus} shows the distribution of the global OP, $\xi^+$ and $\xi^-$, indicative of in-phase and out-of-phase orientation of the \textit{guest} ions across the layers, respectively. $\xi^+$ peaks around 1 about all the axes, indicating a strong orthorhombic-like correlation between \textit{guest} orientation across the layers. On the other hand, $\xi^-$ peaks around zero, indicating the suppression of out-of-phase correlations across layers. The smaller peak at $\xi^-=-1$ correlates well with the relatively infrequent formation of {\bf n}$_3$-like layers. The time evolution of all order parameters are shown in Figs.~S25 and S26.
	
	The behavior of the order parameters discussed above shows that, in the cubic phase, the \textit{guest} ions prefer the orientation of the orthorhombic phase. Furthermore, the molecular plane-normals are dynamically disordered over the three crystallographic axes, indicating that orthorhombic distortions persist in the cubic phase. The latter then emerges as an average over the various equivalent orthorhombic orientations. Hence, we expect the cubic and orthorhombic (Phase-I) phases to have similar free energy. This explains the proximity of the temperatures for transition from Phase-II to Phase-I (predicted) and the cubic phase.
	
	\subsubsection{Nature of \textit{guest} disordering}
	
	One of the primary sources of the internal motion in the organic \textit{guests} is their rotational-librational motion in the cuboctahedral cavity. 
	In general, molecular libration occurs in a timescale of few sub-picoseconds, whereas the molecular rotations exhibit a time constant of a few picoseconds, which further depends on the type of the \textit{guest} involved~\cite{frost2014atomistic,frost2016moving}. It can be expected that the \textit{guest} molecules display complete rotational mobility at high temperatures. For instance, the molecular reorientation at 300 K has time constants of $1.86-3$ ps for MA~\cite{leguy2015dynamics,leguy2016dynamic,maity2022deciphering} and $\approx2$ ps for FA~\cite{weller2015cubic}.
	
	\ch{MHy} being a larger \textit{guest} is expected to have rather slow dynamics due to strong \textit{guest/host} interaction (stronger in Cl than in Br due to higher electronegativity). This is evident with the presence of stronger (weaker) H-bonds between the \textit{guest}, \ch{MHy+} and \ch{Cl-} (\ch{Br-}) due to the smaller (larger) size of the \ch{Cl-} (\ch{Br-}) ion, and the availability of a smaller (larger) cuboctahedral void occupied by the \textit{guest}~\cite{drozdowski2023broadband}. The auto-correlation function (ACF) of the normal to the $i^{\rm th}$ molecular plane ($\mathbf{n}_i$), denoted by $ C_i(t) $, and the average ACF $C(t)$ are defined as follows: 
	\begin{align}\label{eq_ch4:acorr_fn_sep}
		C_i(t)&=\left\langle\left(\mathbf{n}_i(0)\cdot\mathbf{n_i}(t)\right)\right\rangle \\
		C(t)&=\frac{1}{N} \sum_{i=1}^{N} C_i(t)
	\end{align}
	
	\begin{figure}[tbh!]
		\centering
		\includegraphics[width=\columnwidth]{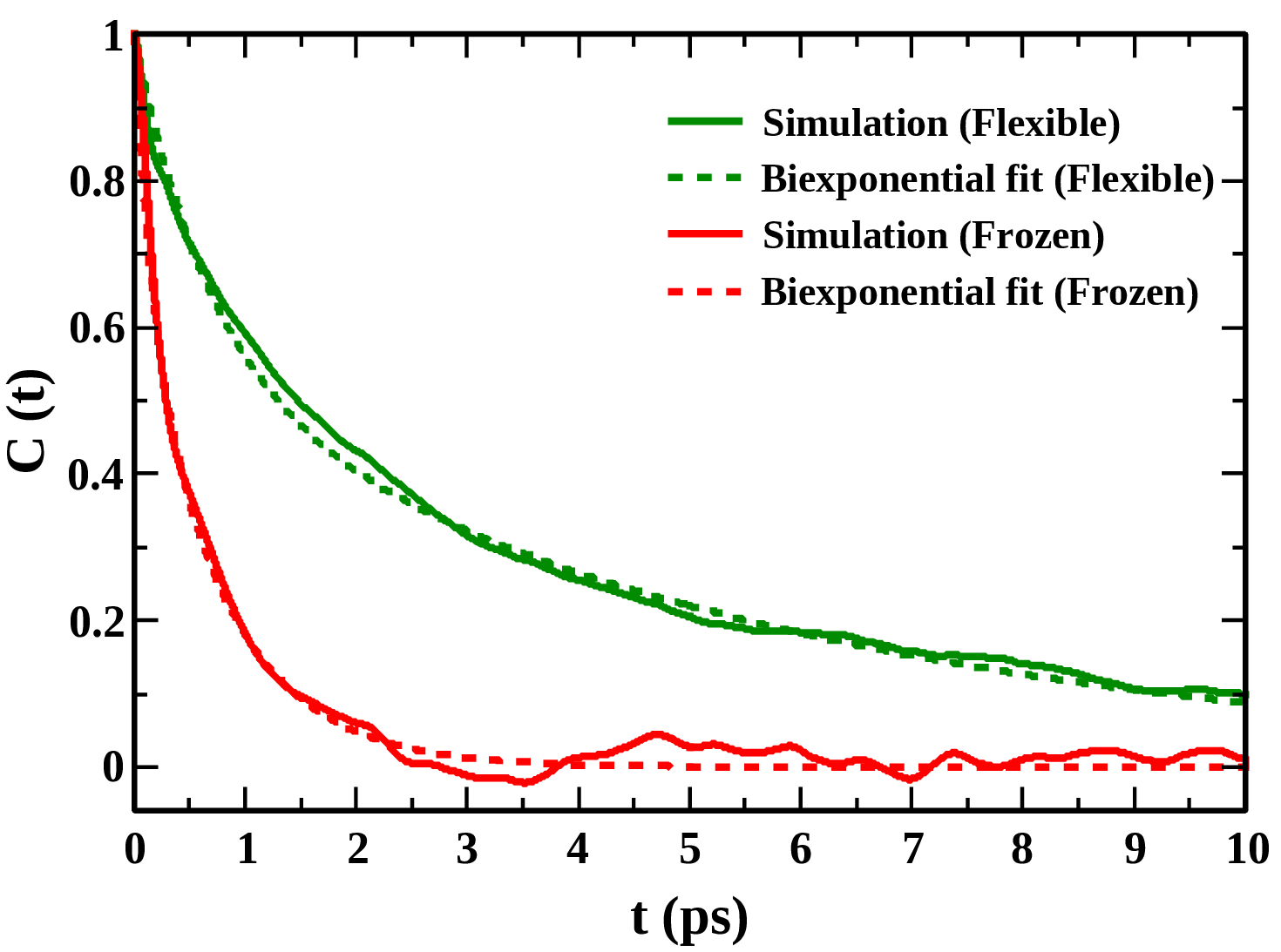}
		\caption{Autocorrelation of the orientation of the normal to the molecular plane (\textbf{$\hat{n}(\theta_n,\phi_n)$)} obtained by fitting the simulation data to the bi-exponential curve. Green curves are for the flexible \textit{host}, and the red ones are for the frozen \textit{host}. Both simulated (shown by solid lines) and calculated data (shown by broken lines) are plotted on the same curve.}
		\label{fig_ch4:acorr_norm_all}
	\end{figure}
	
	\begin{table}
		\caption{Parameters of the bi-exponential fitting of the autocorrelation function in flexible and frozen \textit{host}.}
		\label{tab_ch4tab:biexp_param}
		\centering
		\begin{tabular}{ccc}
			\hline
			\hline
			Parameters for &Flexible & Frozen \\
			biexponential&\textit{host}&\textit{host}\\
			fitting&&\\
			\hline
			\hline
			A & $0.74\pm0.01$&$0.82\pm0.01$\\
			\hline
			$\tau_{jump}$ (ps) & $5.46\pm 0.02$ &$0.76\pm 0.01$\\
			\hline
			$\tau_{fast}$ (ps) & $0.78\pm 0.01$&$0.19\pm 0.01$\\
			\hline
			\hline	
		\end{tabular}
	\end{table}

	The average orientational ACF defined in Eq.~6 is plotted in Fig.~\ref{fig_ch4:acorr_norm_all}. Characteristic time-scales involved are extracted here {\it via} a bi-exponential fit~\cite{liang2023structural} are shown in Table~\ref{tab_ch4tab:biexp_param}. Alternatively, a diffusive model can also be used for the fit, the details of which have been provided in the supplementary, along with the relevant time scales. The ACF confirms a diffusive behavior involving both a fast librational timescale ($\tau_{fast}$) as well as a slow reorientational time-scale ($\tau_{jump}$) over which the molecules jump from one preferred orientation to another. Hence, the molecules get dynamically disordered (also see Fig.~S31). These organic dipoles do not behave like free rotors but are instead disordered over a finite number of orientations, which is a characteristic of their sticky dynamics. This would also be evident later from the strong H- and coordinate bonds formed between this pair of \textit{guest/host}.

	\subsubsection{Energy barriers for \textit{guest} reorientation in \ch{MHyPbBr3} and \ch{MHyPbCl3}}
	
	\begin{figure*}
		\centering
		\subfigure[]{\includegraphics[width=0.28\textwidth]{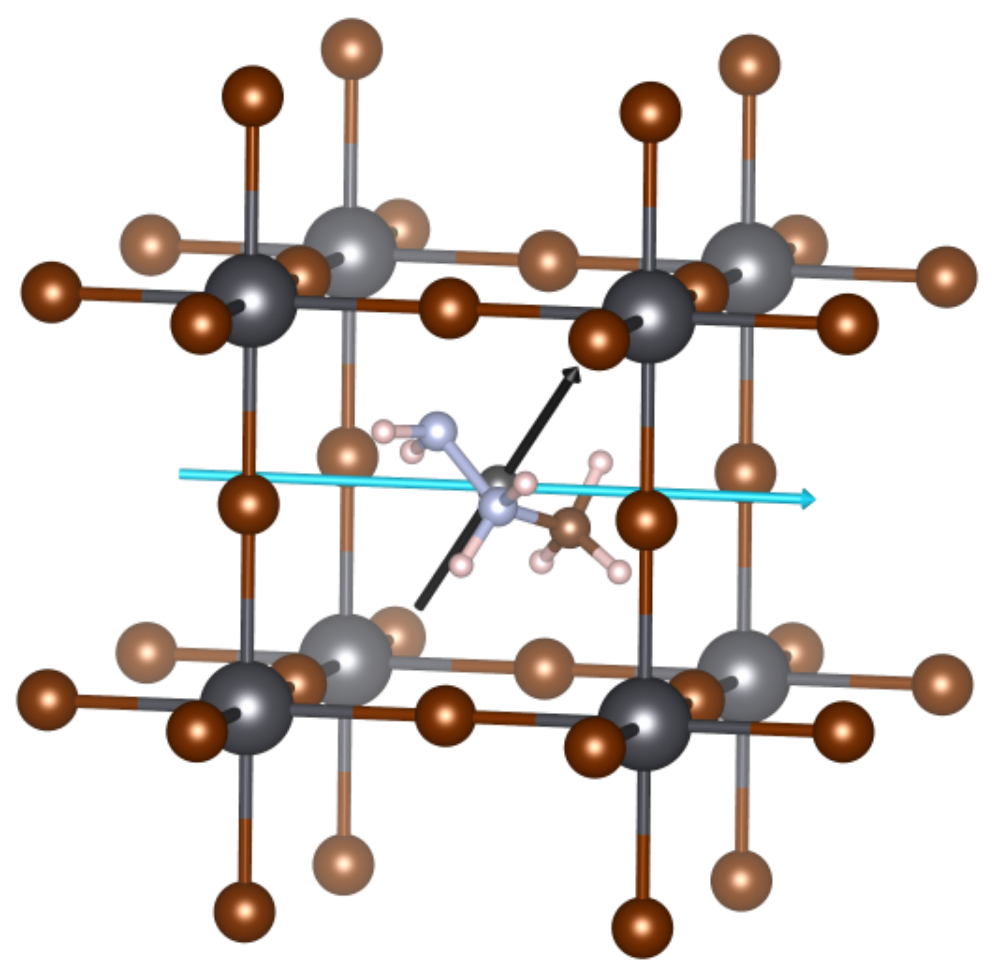}\label{subfig_ch4:fl1_rot_about_axis_ff2}}
		\hfill
		\subfigure[]{\includegraphics[width=0.35\textwidth]{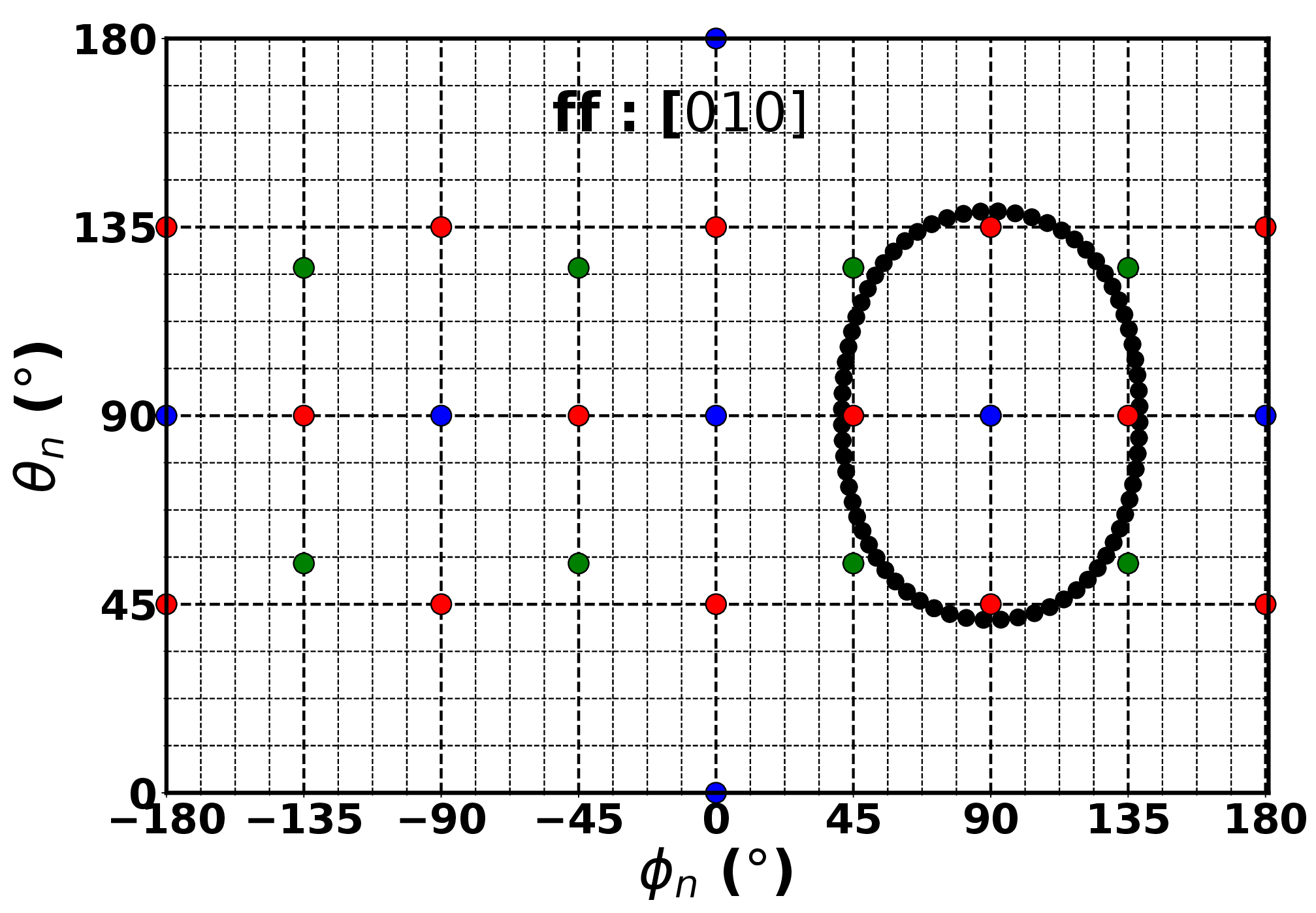}\label{subfig_ch4:fl1_ff2_high_symm_rot_axis}}
		\hfill
		\subfigure[]
		{\includegraphics[width=0.35\textwidth]{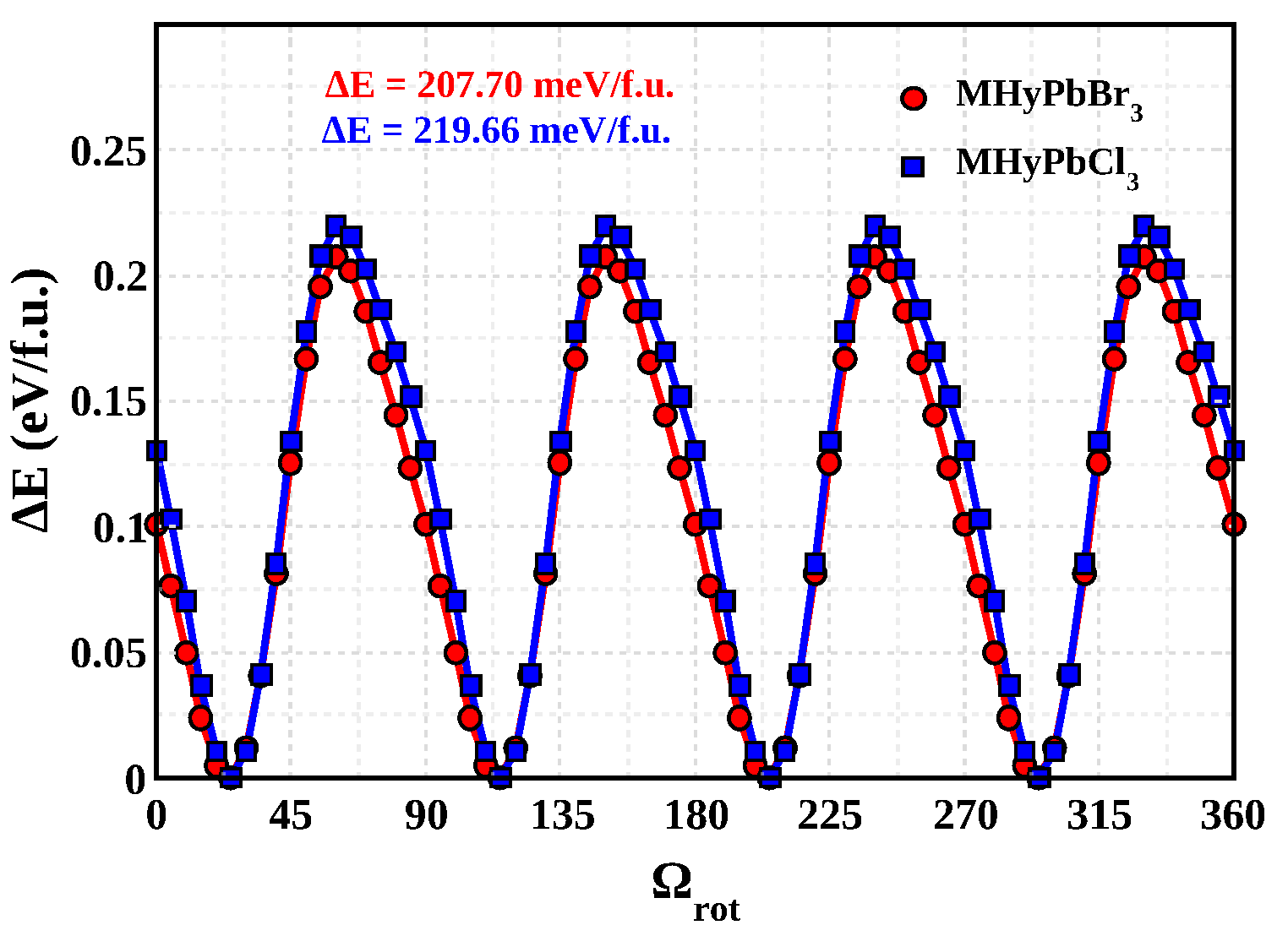}\label{subfig_ch4:rot_barr_fl_cl_br}}
	
		\caption{\label{fig_ch4:high_symm_rot}~\subref{subfig_ch4:fl1_rot_about_axis_ff2}~Schematic of $\mathbf{\hat{n}}$ and rotation axis. The black arrow represents the normal to the molecular plane. Cyan arrow represents the rotation axis, $[01\bar{1}]$ (cubic \textit{ed}).~\subref{subfig_ch4:fl1_ff2_high_symm_rot_axis}~Rotation of $\mathbf{\hat{n}}$ about \textit{ff}-axis $[010]$ to obtain the preferential sites. The colored dots denote the high-symmetry directions in the cubic phase indicated by blue (face to face (\textit{ff})), green (body diagonal (\textit{bd})), and red (edge diagonal (\textit{ed})).~\subref{subfig_ch4:rot_barr_fl_cl_br}~Rotational energy barriers in the cubic phase of \ch{MHyPbBr3} and \ch{MHyPbCl3} at the minimum energy volume and starting from the \textit{guest} orientations for a flexible \textit{host}.}
	\end{figure*}
	
	Since the chloride perovskite is not stable in the cubic phase we did not perform an MD in this structure. However, to probe the origin of the absence of this phase, we compared the rotation barriers between the bromide and chloride at 0K. The DFT-optimized cubic lattice parameters were found to be $5.69~\text{\AA}$ for the chloride and $5.88~\text{\AA}$ for the bromide perovskite (see Fig.~S36(b)). We consider rotating the MHy ion about the [010] axis ($ff$) starting from the orientation labeled as II in Fig.~\ref{subfig_ch4:flex1_norm_orient}. At 300 K, free energy along this path corresponds to the blue curve in Fig.~\ref{subfig_ch4:flex1_free_en_2D}. We performed this rotation for both halide varieties at their respective optimized structures, and the resulting total energy profiles are plotted in Fig.~\ref{subfig_ch4:rot_barr_fl_cl_br}. During rotation, four energy minimum states are encountered, consistent with the symmetry of the \textit{ff}-axis. The barriers separating nearby minima is $\approx 12$ meV/f.u. higher in the chloride. This is equivalent to a temperature difference of around $150$ K. This leads to two important conclusions: (a) the cubic phase of \ch{MHyPbBr3} is kinetically more favorable than \ch{MHyPbCl3} because of smaller energy barriers, and (b) even if a disordered (cubic) state were to exist in Cl, it will appear around $570$ K, which is physically not realizable for this system since it decomposes at around 490 K~\cite{maczka2020three}. The higher transition temperature in the chloride is consistent with the barrier values of rotation about other high symmetry axes as well, reaching almost $1300$ K for the $ed:[01\bar{1}]$-axis (see Sec~SII-A5). The difference in the behavior of the two perovskites can be related to the strength of the HG coupling, which, because of the smaller size and higher electronegativity, is expected to be stronger in Cl. Therefore, higher thermal energy would be required to weaken the coupling for disordering to take place as is reflected in the increase of the phase transition temperature.
	
	\subsubsection{Evolution of \textit{host/guest} interaction}
	
	The HG coupling strength can be measured as a factor of the hydrogen (H-) and coordinate bonds since they are the direct links between the \textit{guest} and the \textit{host}. We plot the radial distribution function (RDF) of these bonds for both the frozen and flexible \textit{host} at 470 K, as shown by Fig.~S34. A very strong coordinate bond is formed between \ch{Pb} and the terminal nitrogen (N2) of the \textit{guest}, which is represented by a clear peak with the bond extending till around $3.85~\text{\AA}$ (Fig.~S34(b)). This is used as the cut-off to decide whether a particular N2$\cdots$Pb pair forms the bond or not. However, for the case of H-bonds (both C-H$\cdots$Br and N-H$\cdots$Br), no distinct peak is observed as such; however, it is more predominant for the frozen \textit{host} and extends up till $3.20~\text{\AA}$ for N-H$\cdots$Br and $3.80~\text{\AA}$ for C-H$\cdots$Br, which are also the cutoffs taken for further calculations (Fig.~S34(a)). While the coordinate bonds seem to persist, the hydrogen bonds have a diffusive nature at this high temperature, indicative of their significant weakening. This is similar to what has been reported for \ch{MAPbBr3}.~\cite{maity2022deciphering} The extent of these bonds matches with the experimentally reported range; hence has been used for obtaining the correlation from both the flexible and frozen \textit{host} simulations. 

\begin{figure*}[tbh!]
		\centering
		\subfigure[]{\includegraphics[width=0.45\textwidth]{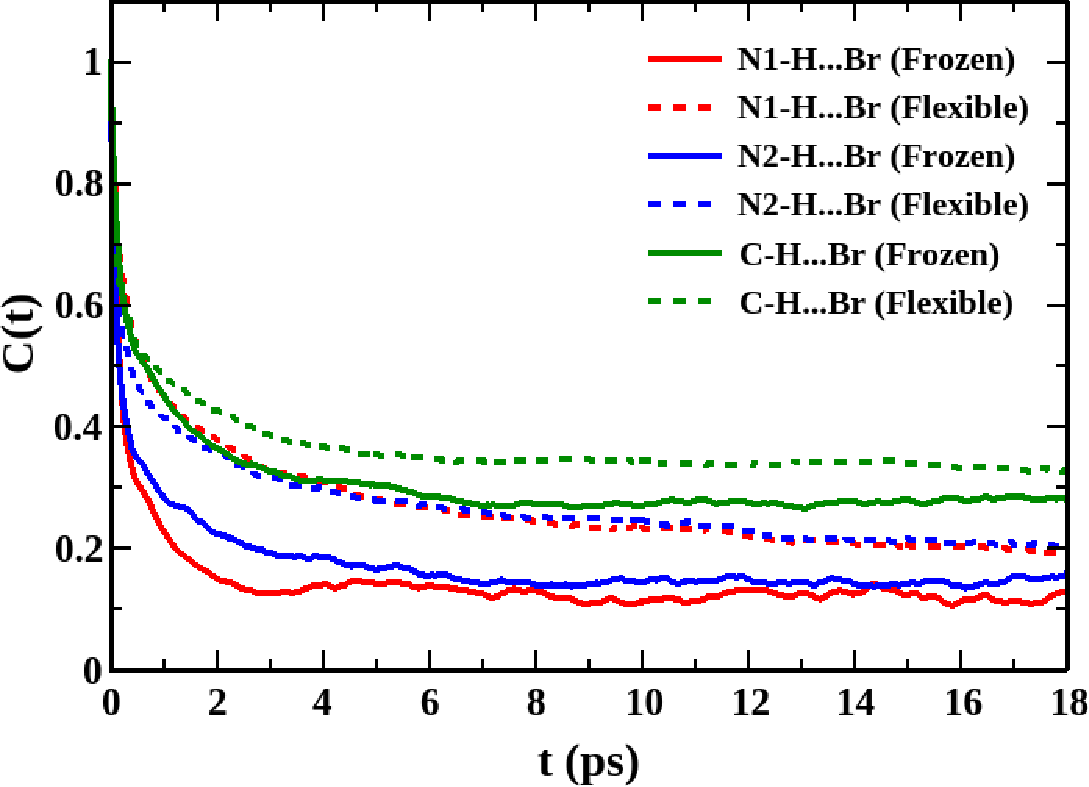}\label{subfig_ch4:h_bond_all}}
		\hfill	
		\subfigure[]{\includegraphics[width=0.45\textwidth]{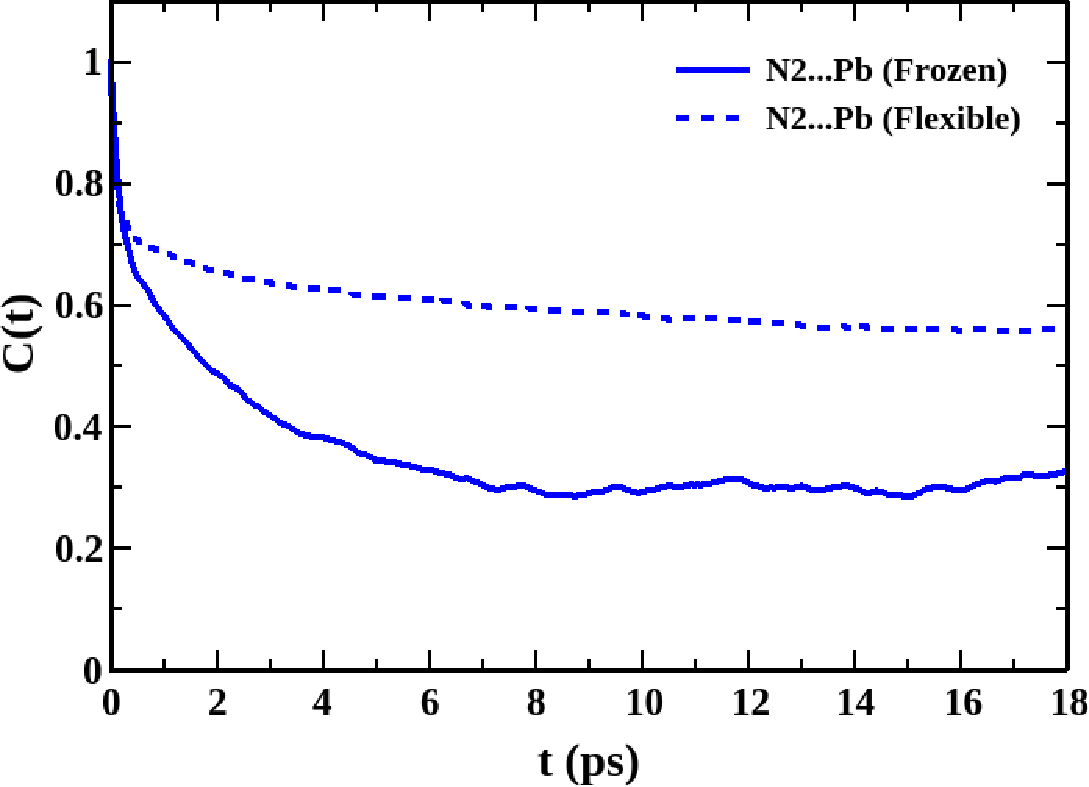}\label{subfig_ch4:corrd_bond_all}}
		
		\caption{\label{fig_ch4:weak_bonds}~Weak H- and coordinate bonds:~\subref{subfig_ch4:h_bond_all}~Autocorrelation of the H-bond under the criteria of $d_{\mathrm{H}\cdots \mathrm{Br}}<3.20~\text{\AA}$. ~\subref{subfig_ch4:corrd_bond_all}~Autocorrelation of the coordinate bond (N2$\cdots$Pb) under the criterion of $d_{\mathrm{N}\cdots \mathrm{Pb}}<3.80~\text{\AA}$.}
	\end{figure*}

	Fig.~\ref{fig_ch4:weak_bonds}\subref{subfig_ch4:h_bond_all} shows the dynamical behavior of the H-bonds as obtained by using the following correlation function~\cite{maity2022deciphering}:
	\begin{equation}\label{eq_ch4:h_bond_acorr}
		C_{X-H\cdots Br}(t)=\frac{1}{N}\sum_{i=1}^{N} \langle \mathrm{\mathbf{H}}_i(0)\cdot \mathrm{\mathbf{H}}_i(t) \rangle
	\end{equation}
	Here $\mathbf{H}(t)=[\delta_{X-H\cdots Br_1},....,\delta_{X-H\cdots Br_{12}}]$, and $\delta_{X-H\cdots Br}$ equals 1 if the $\mathrm{H}\cdots \mathrm{Br}$ distance lies within the above-mentioned cut-offs for each type or 0 elsewhere. The timescale of rotation (fast) of the H-atoms about the molecule lies in the range of $0.3-0.6$ ps and is slower than reported for smaller \textit{guests}~\cite{sharma2020contrasting,bernard2018methylammonium,li2018activation,maity2022deciphering}. It should be noted that the relaxation of the hydrogen bonds is faster for N1 followed closely by N2 while C-H$\cdots$Br relaxes the slowest, which is also evident from the peaks in the RDF plot for the bonds. N$\cdots$Pb relaxes even slower than all the hydrogen bonds and is the strongest \textit{host}/\textit{guest} bonds. The latter is responsible for polarizing the molecule into picking up specific orientations. The sequence of relaxation rates and, hence, the bond strengths remains the same for both the regimes of the \textit{host}. However, the relaxations, hence the timescale of rotations, are slower for the flexible case. Hence, the \textit{guest/host} coupling is strengthened in this case, which leads to a stronger anisotropic \textit{host} environment for the \textit{guest}. It is also important to mention that the stronger N$\cdots$Pb coordinate bond likely anchors the MHy ions in the orthorhombic arrangements even in the cubic phase, thereby also making the formation of {\bf n}$_3$-like phase infrequent.
	
	The parameters for \textit{host} evolution have been explained in Sec.~SII-B1, which confirms the cubic nature of the \ch{PbBr6} octahedra.
	From all of the above analyses, we can conclude that the cubic phase of \ch{MHyPbBr3} is dynamically disordered, the evidence of which has been reported for both the sub-lattices, the \textit{guest} as well as the \textit{host} in addition to local static distortion of the \textit{host}. Also, \ch{MHy+}, being a larger organic molecule, shows quite slow dynamics due to the stronger hydrogen and coordinate bonds in comparison to other halide analogs. These stable bonds must be responsible for strong \textit{guest/host} coupling effects in the compounds employing \ch{MHy} as the \textit{guest} ion.
	
	\section{Conclusion}
	We have investigated the origin of the missing phases during temperature-driven transitions in two halide analogs of an over-tolerant hybrid perovskite \ch{MHyPbX3}. In particular, we were intrigued by the lack of an intermediate ordered phase in \ch{MHyPbBr3} as well as a missing disordered cubic phase in \ch{MHyPbCl3}. To this end, we computed the relative free energies of the ordered phases (Phase-I and Phase-II) in the two compounds under the quasi-harmonic approximation. We found an unstable normal mode in the experimental Phase-I structure of the chloride, along which dynamically stable structures can be achieved by displacement. The same was also found to be true in the analogous structure of the bromide. The shallow double well potential underlying these unstable modes was included in the computation of the free energies through a quartic form, which improved the prediction of transition temperatures between the two ordered phases. The predicted temperature in chloride agreed well with experiments. In the bromide, our predictions indicate an, in principle, stable Phase-I occurring at a temperature of only 10 K above the cubic phase. This prompted us to look closely at the cubic phase at 300 K using {\it ab initio} molecular dynamics simulations. Using an order parameter that defines the collective orientation of the \textit{guest} MHy ions, we show that the experimental cubic phase can result out of disordering among locally orthorhombic structures. This also justifies the similar transition temperatures to either phase. At 300 K, the \ch{N-H...Br} and \ch{C-H...Br} hydrogen bonds are sufficiently weakened that the MHy ions can rotate freely about the \ch{N...Pb} coordinate bonds. The latter remains relatively stronger even at 300 K, tethering the MHy ions to the same central Pb layer. This is also the origin of the strong \textit{host/guest} coupling in the system. The collective motion of all these tethered MHy ions leads to disordering of the molecular planes over the three crystallographic planes. Detachment of the MHy ions from this and reattachment to another Pb layer is infrequent and leads to an equivalent orthorhombic ({\bf n}$_3$-like) arrangement.
	
	We also find that the more polar Phase-I in both perovskites could be rendered dynamically stable by the application of pressure. Indeed, experimentally reported pressure-induced phase in the bromide~\cite{mao2023pressure,szafranski2024structural} resembles the Phase-I structure locally, offering support to our claim.
	
	AIMD simulations on the cubic phase of \ch{MHyPbBr3} allowed us to establish not only the preferred orientations of the MHy ions but also provided insights into the nature and strength of the \textit{host/guest} coupling in this over-tolerant perovskite. Assuming similar orientations to also be preferred in a hypothetical cubic phase of \ch{MHyPbCl3}, we compared the energy barrier for rotation of the \textit{guest} in the two halides. This revealed that a disordering transition in the chloride would require 150 K higher than that in the bromide. However, since the chloride is known to decompose at 490~K, the transition to a disordered cubic phase is not observed in the compound. 
	
	We also propose a useful model to assess the sequence of temperature-driven transitions in \textit{host/guest} systems. These criteria, when applied to \ch{MHyPbX3}, predict sequences in agreement with the experiment as well as those demonstrated in our DFT calculations. Similar arguments could be extended to other OIHPs to engineer relevant couplings and stabilize phases of interest at room temperature.
	
	\begin{acknowledgements}
		The authors gratefully acknowledge computational resources provided by IISER Bhopal as well as “PARAM Shivay” at Indian Institute of Technology (BHU), Varanasi, which is implemented by C-DAC and supported by the Ministry of Electronics and Information Technology (MeitY) and the Department of Science and Technology (DST), Government of India, through National Supercomputing Mission (NSM). P.S. acknowledges funding through UGC-JRF (India) for carrying out the Ph.D. program. S.M. acknowledges funding through the Integrated Ph.D. program at IISER Bhopal. 
	\end{acknowledgements}

	\bibliographystyle{unsrt}
	\bibliography{mhpx_bib}

\begin{thebibliography}{10}

\bibitem{stranks2015metal}
Samuel~D Stranks and Henry~J Snaith.
\newblock Metal-halide perovskites for photovoltaic and light-emitting devices.
\newblock {\em Nature Nanotechnology}, 10(5):391--402, 2015.

\bibitem{saparov2016organic}
Bayrammurad Saparov and David~B Mitzi.
\newblock Organic--inorganic perovskites: structural versatility for functional
  materials design.
\newblock {\em Chemical Reviews}, 116(7):4558--4596, 2016.

\bibitem{babayigit2016ethirajan}
A~Babayigit, Ethirajan A., Muller M., and Conings B.
\newblock Toxicity of organometal halide perovskite solar cells.
\newblock {\em Nature Materials}, 15:247--251, 2016.

\bibitem{vogt1993high}
T~Vogt and Wolfgang~W Schmahl.
\newblock The high-temperature phase transition in perovskite.
\newblock {\em Europhysics Letters}, 24(4):281, 1993.

\bibitem{dittrich2015temperature}
Thomas Dittrich, Celline Awino, Pongthep Prajongtat, Bernd Rech, and Martha~Ch
  Lux-Steiner.
\newblock Temperature dependence of the band gap of ch3nh3pbi3 stabilized with
  pmma: a modulated surface photovoltage study.
\newblock {\em The Journal of Physical Chemistry C}, 119(42):23968--23972,
  2015.

\bibitem{herz2016charge}
Laura~M Herz.
\newblock Charge-carrier dynamics in organic-inorganic metal halide
  perovskites.
\newblock {\em Annual Review of Physical Chemistry}, 67:65--89, 2016.

\bibitem{maczka2020three}
M.~M\k{a}czka, A.~G\k{a}gor, J.~K. Zar\k{e}ba, D.~Stefa{\'n}ska, M.~Drozd,
  S.~Balciunas, \v{S}im\.enas, J.~M., Banys, and A.~Sieradzki.
\newblock Three-dimensional perovskite methylhydrazinium lead chloride with two
  polar phases and unusual second-harmonic generation bistability above room
  temperature.
\newblock {\em Chemistry of Materials}, 32(9):4072--4082, 2020.

\bibitem{maczka2020methylhydrazinium}
Miros{\l}aw Maczka, Maciej Ptak, Anna G\c{a}gor, Dagmara Stefa{\'n}ska, Jan~K
  Zar\c{e}ba, and Adam Sieradzki.
\newblock Methylhydrazinium lead bromide: noncentrosymmetric three-dimensional
  perovskite with exceptionally large framework distortion and green
  photoluminescence.
\newblock {\em Chemistry of Materials}, 32(4):1667--1673, 2020.

\bibitem{drozdowski2022three}
Dawid Drozdowski, Anna G\k{a}gor, Dagmara Stefa{\'n}ska, Jan~K Zar\k{e}ba,
  Katarzyna Fedoruk, Miros{\l}aw M\k{a}czka, and Adam Sieradzki.
\newblock Three-dimensional methylhydrazinium lead halide perovskites:
  Structural changes and effects on dielectric, linear, and nonlinear optical
  properties entailed by the halide tuning.
\newblock {\em The Journal of Physical Chemistry C}, 126(3):1600--1610, 2022.

\bibitem{pradhi_mhpc}
Pradhi Srivastava, Sayan Maity, and Varadharajan Srinivasan.
\newblock Guest-induced phase transition leads to polarization enhancement in
  \ch{MHyPbCl3}.
\newblock {\em arXiv}, 2023.

\bibitem{svane2017strong}
Katrine~L Svane, Alexander~C Forse, Clare~P Grey, Gregor Kieslich, Anthony~K
  Cheetham, Aron Walsh, and Keith~T Butler.
\newblock How strong is the hydrogen bond in hybrid perovskites?
\newblock {\em The Journal of Physical Chemistry Letters}, 8(24):6154--6159,
  2017.

\bibitem{giannozzi2009quantum}
Paolo Giannozzi, Stefano Baroni, Nicola Bonini, Matteo Calandra, Roberto Car,
  Carlo Cavazzoni, Davide Ceresoli, Guido~L Chiarotti, Matteo Cococcioni, and
  Ismaila Dabo.
\newblock Quantum espresso: a modular and open-source software project for
  quantum simulations of materials.
\newblock {\em Journal of physics: Condensed matter}, 21(39):395502, 2009.

\bibitem{giannozzi2017advanced}
Paolo Giannozzi, Oliviero Andreussi, Thomas Brumme, Oana Bunau, M~Buongiorno
  Nardelli, Matteo Calandra, Roberto Car, Carlo Cavazzoni, Davide Ceresoli, and
  Matteo Cococcioni.
\newblock Advanced capabilities for materials modelling with quantum espresso.
\newblock {\em Journal of physics: Condensed matter}, 29(46):465901, 2017.

\bibitem{rappe1990optimized}
Andrew~M Rappe, Karin~M Rabe, Efthimios Kaxiras, and JD~Joannopoulos.
\newblock Optimized pseudopotentials.
\newblock {\em Physical Review B}, 41(2):1227, 1990.

\bibitem{vanderbilt1990soft}
David Vanderbilt.
\newblock Soft self-consistent pseudopotentials in a generalized eigenvalue
  formalism.
\newblock {\em Physical Review B}, 41(11):7892, 1990.

\bibitem{perdew1996generalized}
John~P Perdew, Kieron Burke, and Matthias Ernzerhof.
\newblock Generalized gradient approximation made simple.
\newblock {\em Physical Review Letters}, 77(18):3865, 1996.

\bibitem{perdew2008restoring}
John~P Perdew, Adrienn Ruzsinszky, G{\'a}bor~I Csonka, Oleg~A Vydrov, Gustavo~E
  Scuseria, Lucian~A Constantin, Xiaolan Zhou, and Kieron Burke.
\newblock Restoring the density-gradient expansion for exchange in solids and
  surfaces.
\newblock {\em Physical Review Letters}, 100(13):136406, 2008.

\bibitem{grimme2006semiempirical}
Stefan Grimme.
\newblock Semiempirical gga-type density functional constructed with a
  long-range dispersion correction.
\newblock {\em Journal of Computational Chemistry}, 27(15):1787--1799, 2006.

\bibitem{barone2009role}
Vincenzo Barone, Maurizio Casarin, Daniel Forrer, Michele Pavone, Mauro Sambi,
  and Andrea Vittadini.
\newblock Role and effective treatment of dispersive forces in materials:
  Polyethylene and graphite crystals as test cases.
\newblock {\em Journal of Computational Chemistry}, 30(6):934--939, 2009.

\bibitem{lee2010higher}
Kyuho Lee, {\'E}amonn~D Murray, Lingzhu Kong, Bengt~I Lundqvist, and David~C
  Langreth.
\newblock Higher-accuracy van der waals density functional.
\newblock {\em Physical Review B}, 82(8):081101, 2010.

\bibitem{egger2014role}
David~A Egger and Leeor Kronik.
\newblock Role of dispersive interactions in determining structural properties
  of organic--inorganic halide perovskites: insights from first-principles
  calculations.
\newblock {\em The Journal of Physical Chemistry Letters}, 5(15):2728--2733,
  2014.

\bibitem{wang2014density}
Yun Wang, Tim Gould, John~F Dobson, Haimin Zhang, Huagui Yang, Xiangdong Yao,
  and Huijun Zhao.
\newblock Density functional theory analysis of structural and electronic
  properties of orthorhombic perovskite ch 3 nh 3 pbi 3.
\newblock {\em Physical Chemistry Chemical Physics}, 16(4):1424--1429, 2014.

\bibitem{beck2019structure}
Hubert Beck, Christian Gehrmann, and David~A Egger.
\newblock Structure and binding in halide perovskites: Analysis of static and
  dynamic effects from dispersion-corrected density functional theory.
\newblock {\em APL Materials}, 7(2), 2019.

\bibitem{kresse1995ab}
G~Kresse, J~Furthm{\"u}ller, and J~Hafner.
\newblock Ab initio force constant approach to phonon dispersion relations of
  diamond and graphite.
\newblock {\em Europhysics Letters}, 32(9):729, 1995.

\bibitem{parlinski1997first}
K~Parlinski, ZQ~Li, and Y~Kawazoe.
\newblock First-principles determination of the soft mode in cubic zro 2.
\newblock {\em Physical Review Letters}, 78(21):4063, 1997.

\bibitem{chaput2011phonon}
Laurent Chaput, Atsushi Togo, Isao Tanaka, and Gilles Hug.
\newblock Phonon-phonon interactions in transition metals.
\newblock {\em Physical Review B}, 84(9):094302, 2011.

\bibitem{togo2008first}
Atsushi Togo, Fumiyasu Oba, and Isao Tanaka.
\newblock First-principles calculations of the ferroelastic transition between
  rutile-type and cacl 2-type sio 2 at high pressures.
\newblock {\em Physical Review B}, 78(13):134106, 2008.

\bibitem{phonopy}
A~Togo and I~Tanaka.
\newblock First principles phonon calculations in materials science.
\newblock {\em Scr. Mater.}, 108:1--5, Nov 2015.

\bibitem{wallace1972thermodynamics}
Duane~C Wallace and Herbert Callen.
\newblock Thermodynamics of crystals.
\newblock {\em American Journal of Physics}, 40(11):1718--1719, 1972.

\bibitem{born1985quantentheorie}
Max Born and W~Heisenberg.
\newblock Zur quantentheorie der molekeln.
\newblock {\em Original Scientific Papers Wissenschaftliche Originalarbeiten},
  pages 216--246, 1985.

\bibitem{born2000quantum}
Max Born and Robert Oppenheimer.
\newblock On the quantum theory of molecules.
\newblock In {\em Quantum Chemistry: Classic Scientific Papers}, pages 1--24.
  World Scientific, 2000.

\bibitem{barnett1993born}
Robert~N Barnett and Uzi Landman.
\newblock Born-oppenheimer molecular-dynamics simulations of finite systems:
  Structure and dynamics of (h 2 o) 2.
\newblock {\em Physical Review B}, 48(4):2081, 1993.

\bibitem{kresse1993ab}
Georg Kresse and J{\"u}rgen Hafner.
\newblock Ab initio molecular dynamics for liquid metals.
\newblock {\em Physical Review B}, 47(1):558, 1993.

\bibitem{kresse1996efficiency}
Georg Kresse and J{\"u}rgen Furthm{\"u}ller.
\newblock Efficiency of ab-initio total energy calculations for metals and
  semiconductors using a plane-wave basis set.
\newblock {\em Computational Materials Science}, 6(1):15--50, 1996.

\bibitem{kresse1996efficient}
Georg Kresse and J{\"u}rgen Furthm{\"u}ller.
\newblock Efficient iterative schemes for ab initio total-energy calculations
  using a plane-wave basis set.
\newblock {\em Physical Review B}, 54(16):11169, 1996.

\bibitem{nose1984unified}
Shuichi Nos{\'e}.
\newblock A unified formulation of the constant temperature molecular dynamics
  methods.
\newblock {\em The Journal of Chemical Physics}, 81(1):511--519, 1984.

\bibitem{hoover1985canonical}
William~G Hoover.
\newblock Canonical dynamics: Equilibrium phase-space distributions.
\newblock {\em Physical Review A}, 31(3):1695, 1985.

\bibitem{kresse1994ab}
Georg Kresse and J{\"u}rgen Hafner.
\newblock Ab initio molecular-dynamics simulation of the
  liquid-metal--amorphous-semiconductor transition in germanium.
\newblock {\em Physical Review B}, 49(20):14251, 1994.

\bibitem{kresse1999ultrasoft}
Georg Kresse and Daniel Joubert.
\newblock From ultrasoft pseudopotentials to the projector augmented-wave
  method.
\newblock {\em Physical Review B}, 59(3):1758, 1999.

\bibitem{saidi2016nature}
Wissam~A Saidi and Joshua~J Choi.
\newblock Nature of the cubic to tetragonal phase transition in methylammonium
  lead iodide perovskite.
\newblock {\em The Journal of Chemical Physics}, 145(14):144702, 2016.

\bibitem{lohaus2020thermodynamic}
Stefan~H Lohaus, Michel~B Johnson, Peter~F Ahnn, Claire~N Saunders, Hillary~L
  Smith, Mary~Anne White, and Brent Fultz.
\newblock Thermodynamic stability and contributions to the gibbs free energy of
  nanocrystalline ni 3 fe.
\newblock {\em Physical Review Materials}, 4(8):086002, 2020.

\bibitem{zhang2018intrinsic}
Yue-Yu Zhang, Shiyou Chen, Peng Xu, Hongjun Xiang, Xin-Gao Gong, Aron Walsh,
  and Su-Huai Wei.
\newblock Intrinsic instability of the hybrid halide perovskite semiconductor
  ch3nh3pbi3.
\newblock {\em Chinese Physics Letters}, 35(3):036104, 2018.

\bibitem{baroni2001phonons}
Stefano Baroni, Stefano De~Gironcoli, Andrea Dal~Corso, and Paolo Giannozzi.
\newblock Phonons and related crystal properties from density-functional
  perturbation theory.
\newblock {\em Reviews of modern Physics}, 73(2):515, 2001.

\bibitem{maity2024cooperative}
Sayan Maity, Suraj Verma, Lavanya~M Ramaniah, and Varadharajan Srinivasan.
\newblock Cooperative octahedral tilt modes drive coexisting displacive and
  order--disorder pressure-induced phase transitions in mapbbr3.
\newblock {\em The Journal of Physical Chemistry C}, 128(20):8531–8539, 2024.

\bibitem{frost2014atomistic}
Jarvist~M Frost, Keith~T Butler, Federico Brivio, Christopher~H Hendon, Mark
  Van~Schilfgaarde, and Aron Walsh.
\newblock Atomistic origins of high-performance in hybrid halide perovskite
  solar cells.
\newblock {\em Nano Letters}, 14(5):2584--2590, 2014.

\bibitem{frost2016moving}
Jarvist~M Frost and Aron Walsh.
\newblock What is moving in hybrid halide perovskite solar cells?
\newblock {\em Accounts of Chemical Research}, 49(3):528--535, 2016.

\bibitem{leguy2015dynamics}
Aurelien~MA Leguy, Jarvist~Moore Frost, Andrew~P McMahon, Victoria~Garcia
  Sakai, W~Kockelmann, ChunHung Law, Xiaoe Li, Fabrizia Foglia, Aron Walsh, and
  Brian~C O’regan.
\newblock The dynamics of methylammonium ions in hybrid organic--inorganic
  perovskite solar cells.
\newblock {\em Nature Communications}, 6(1):7124, 2015.

\bibitem{leguy2016dynamic}
Aur{\'e}lien~MA Leguy, Alejandro~R Go{\~n}i, Jarvist~M Frost, Jonathan Skelton,
  Federico Brivio, Xabier Rodr{\'\i}guez-Mart{\'\i}nez, Oliver~J Weber,
  Anuradha Pallipurath, M~Isabel Alonso, and Mariano Campoy-Quiles.
\newblock Dynamic disorder, phonon lifetimes, and the assignment of modes to
  the vibrational spectra of methylammonium lead halide perovskites.
\newblock {\em Physical Chemistry Chemical Physics}, 18(39):27051--27066, 2016.

\bibitem{maity2022deciphering}
Sayan Maity, Suraj Verma, Lavanya~M Ramaniah, and Varadharajan Srinivasan.
\newblock Deciphering the nature of temperature-induced phases of mapbbr3 by ab
  initio molecular dynamics.
\newblock {\em Chemistry of Materials}, 2022.

\bibitem{weller2015cubic}
Mark~T Weller, Oliver~J Weber, Jarvist~M Frost, and Aron Walsh.
\newblock Cubic perovskite structure of black formamidinium lead iodide,
  $\alpha$-[hc (nh2) 2] pbi3, at 298 k.
\newblock {\em The Journal of Physical Chemistry Letters}, 6(16):3209--3212,
  2015.

\bibitem{drozdowski2023broadband}
Dawid Drozdowski, Katarzyna Fedoruk, Adam Kabanski, Miros{\l}aw Maczka, Adam
  Sieradzki, and Anna Gagor.
\newblock Broadband yellow and white emission from large octahedral tilting in
  (110)-oriented layered perovskites: imidazolium-methylhydrazinium lead
  halides.
\newblock {\em Journal of Materials Chemistry C}, 11(14):4907--4915, 2023.

\bibitem{liang2023structural}
Xia Liang, Johan Klarbring, William~J Baldwin, Zhenzhu Li, G{\'a}bor
  Cs{\'a}nyi, and Aron Walsh.
\newblock Structural dynamics descriptors for metal halide perovskites.
\newblock {\em The Journal of Physical Chemistry C}, 127(38):19141--19151,
  2023.

\bibitem{sharma2020contrasting}
VK~Sharma, R~Mukhopadhyay, A~Mohanty, M~Tyagi, JP~Embs, and DD~Sarma.
\newblock Contrasting behaviors of fa and ma cations in a pbbr3.
\newblock {\em The Journal of Physical Chemistry Letters}, 11(22):9669--9679,
  2020.

\bibitem{bernard2018methylammonium}
Guy~M Bernard, Roderick~E Wasylishen, Christopher~I Ratcliffe, Victor Terskikh,
  Qichao Wu, Jillian~M Buriak, and Tate Hauger.
\newblock Methylammonium cation dynamics in methylammonium lead halide
  perovskites: a solid-state nmr perspective.
\newblock {\em The Journal of Physical Chemistry A}, 122(6):1560--1573, 2018.

\bibitem{li2018activation}
Jingrui Li, Mathilde Bouchard, Peter Reiss, Dmitry Aldakov, St\'ephanie Pouget,
  Renaud Demadrille, Cyril Aumaitre, Bernhard Frick, David Djurado, and Mariana
  Rossi.
\newblock Activation energy of organic cation rotation in ch3nh3pbi3 and
  cd3nh3pbi3: quasi-elastic neutron scattering measurements and
  first-principles analysis including nuclear quantum effects.
\newblock {\em The Journal of Physical Chemistry Letters}, 9(14):3969--3977,
  2018.

\bibitem{mao2023pressure}
Yuhong Mao, Songhao Guo, Xu~Huang, Kejun Bu, Zhongyang Li, Phuong Q.~H. Nguyen,
  Gang Liu, Qingyang Hu, Dongzhou Zhang, Yongping Fu, Wenge Yang, and Xujie
  Lü.
\newblock Pressure-modulated anomalous organic–inorganic interactions enhance
  structural distortion and second-harmonic generation in mhypbbr3 perovskite.
\newblock {\em Journal of the American Chemical Society}, 145(43):23842--23848,
  2023.

\bibitem{szafranski2024structural}
Marek Szafra{\'n}ski.
\newblock Structural and optical properties of methylhydrazinium lead bromide
  perovskites under pressure.
\newblock {\em Journal of Materials Chemistry A}, 12(0):2391--2399, 2024.

\end{thebibliography}
\end{document}